\input ./style/arxiv-general.cfg
\documentclass[MSNbibl,nameyear,dvips]{arxstspdf}
\makeatletter
   \@ifpackageloaded{graphicx}{}{\usepackage{graphicx}}
\makeatother
\usepackage{flushend}
\usepackage{stfloats}

\volume{30}
\issue{2}
\pubyear{2015}
\firstpage{147}
\lastpage{163}
\doi{10.1214/14-STS487} % kopijuoti is 'New paper accepted'
\docsubty{FLA}

\makeatletter
\def\wh{\widehat}
\def\var{\operatorname{var}}
\def\cov{\operatorname{cov}}
\def\wh{\widehat}
\newcommand{\R}{\mathbb{R}}
\newcommand{\Sp}{\mathbb{S}}
\newcommand{\bA}{\mathbf{A}}
\newcommand{\bC}{\mathbf{C}}
\newcommand{\bZ}{\mathbf{Z}}
\newcommand{\bh}{\mathbf{h}}
\newcommand{\bff}{\mathbf{f}}
\newcommand{\bk}{\mathbf{k}}
\newcommand{\by}{\mathbf{y}}
\newcommand{\bs}{\mathbf{s}}
\newcommand{\bv}{\mathbf{v}}
\newcommand{\bx}{\mathbf{x}}
\newcommand{\bmu}{\boldsymbol{\mu}}
\newcommand{\blambda}{\boldsymbol{\lambda}}
\newcommand{\bomega}{\boldsymbol{\omega}}
\newcommand{\bgamma}{\boldsymbol{\gamma}}
\newcommand{\bxi}{\boldsymbol{\xi}}
\newcommand{\bSigma}{\boldsymbol{\Sigma}}
\newcommand{\0}{\mathbf{0}}

\makeatother

\begin{document}
\begin{frontmatter}

\title{Cross-Covariance Functions for Multivariate Geostatistics\thanksref{T1}}%
%\thanksref{T1}
% kai straipsnis turi susijusiu diskusiju ir rejoinder'iu
\relateddois{T1}{Discussed in \relateddoi{d}{10.1214/14-STS515}, \relateddoi{d}{10.1214/14-STS516},
\relateddoi{d}{10.1214/14-STS517} and \relateddoi{d}{10.1214/14-STS518};
rejoinder at \relateddoi{r}{10.1214/14-STS519}.}
\runtitle{Cross-Covariance Functions}
%\pdftitle{}

\begin{aug}
\author[a]{\fnms{Marc G.}~\snm{Genton}\corref{}\ead[label=e1]{marc.genton@kaust.edu.sa}}
\and
\author[b]{\fnms{William}~\snm{Kleiber}\ead[label=e2]{william.kleiber@colorado.edu}}
\runauthor{M. G. Genton and W. Kleiber}
%\pdfauthor{}

\affiliation{King Abdullah University of Science and Technology and
University of Colorado}

\address[a]{\fontsize{9.6}{11.6}{\selectfont Marc G. Genton is Professor, CEMSE Division, King Abdullah
University of Science and Technology,
Thuwal 23955-6900, Saudi Arabia \printead{e1}.}}
\address[b]{\fontsize{9.6}{11.6}{\selectfont William Kleiber is Assistant Professor, Department of
Applied Mathematics, University of Colorado,
Boulder, Colorado 80309-0526, USA \printead{e2}.}}
\end{aug}

% ABSTRACT
%
\begin{abstract}
Continuously indexed datasets with multiple variables have become
ubiquitous in the
geophysical, ecological, environmental and climate sciences,
and pose substantial analysis challenges to scientists and statisticians.
For many years, scientists developed models that aimed at capturing
the spatial behavior for an individual process; only within the last
few decades
has it become commonplace to model multiple processes jointly.
The key difficulty is in specifying the cross-covariance function,
that is, the function responsible for the relationship between distinct
variables. Indeed, these cross-covariance functions must be chosen to be
consistent with marginal covariance functions in such a way that the second-order structure always yields a nonnegative definite covariance matrix.
We review the main approaches to building cross-covariance models, including
the linear model of coregionalization, convolution methods, the multivariate
Mat\'ern and nonstationary and space--time extensions of these among others.
We additionally cover specialized constructions, including those designed
for asymmetry, compact support and spherical domains, with a review of
physics-constrained models.
We illustrate select models on a bivariate regional climate model
output example
for temperature and pressure, along with a bivariate minimum and maximum
temperature observational dataset;
we compare models by likelihood value as well as via
cross-validation co-kriging
studies. The article closes with a discussion of unsolved problems.
\end{abstract}

% KEYWORDS
\end{frontmatter}

%s1 #&#
\section{Introduction}\label{sec:intro}
%%%%%%%%%%%%%%%%%%%%%%%%%%%%%%%%%%%%%%%%%%%%%%%%%%%%%%%%%%%%%%%%%%%%%%%%

%%%%%%%%%%%%%%%%%%%%%%%%%%%%%%%%%%%%%%%%%%%%%%%%%%%%%%%%%%%%%%%%%%%%%%%%
%s1.1 #&#
\subsection{Motivation}

The occurrence of multivariate data indexed by spatial coordinates in a large
number of applications has prompted sustained interest in statistics in recent
years. For instance, in environmental and climate sciences, monitors collect
information on multiple variables
such as temperature, pressure, wind speed and direction and various pollutants.
Similarly, the output of climate models generate
multiple variables, and there are multiple distinct climate models. Physical
models in computer
experiments often involve multiple processes that are indexed by not
only space
and time, but also parameter settings. With the increasing availability and
scientific interest in multivariate processes, statistical science
faces new
challenges and an expanding horizon of opportunities for future exploration.

Geostatistical applications mainly focus on interpolation, simulation
or statistical
modeling. Interpolation or smoothing in spatial statistics usually is
synonymous with kriging, the best linear unbiased prediction under
squared loss
\citep{Cre93}. With multiple variables, interpolation becomes a multivariate
problem, and is traditionally accommodated via co-kriging, the\vadjust{\goodbreak}
multivariate extension of
kriging. Co-kriging is often particularly useful when one variable is of
primary importance, but is correlated with other types of processes
that are
more readily observed (\cite{AlmJou94}; \cite{Wac94}; \cite{Jou99}; \cite{ShmJou99}; \cite{SubPan08}).
Much expository work has been developed on co-kriging,
see \citeauthor{Mye82} (\citeyear{Mye82}, \citeyear{Mye83}, \citeyear{Mye91}, \citeyear{Mye92}), \citet{LonMye97}, \citet{FurGen11}
and \citet{SanJunHua11} for discussion and technical details.

Consider a $p$-dimensional multivariate random field
$\bZ(\bs)=\{Z_1(\bs),\ldots,Z_p(\bs)\}^{\mathrm{T}}$
defined on $\R^d$, $d\geq1$,
where $Z_i(\bs)$ is the $i$th process at location $\bs$, for
$i=1,\ldots,p$.
If $\bZ(\bs)$ is assumed to be a Gaussian multivariate random field,
then only its mean vector $\bmu(\bs)=\mathrm{E}\{\bZ(\bs)\}$ and
cross-covariance matrix function $\bC(\bs_1,\bs_2)=\cov\{\bZ(\bs
_1),\bZ(\bs_2)\}=\{C_{ij}(\bs_1,\bs_2)\}_{i,j=1}^p$ composed of functions
%
%
%e1 #&#
\begin{equation}
\label{ccdef}\hspace*{13pt} C_{ij}(\bs_1,\bs_2)=\cov \bigl
\{Z_i(\bs_1),Z_j(\bs_2) \bigr
\},\quad  \bs_1,\bs_2 \in\R^d,
\end{equation}
for $i,j=1,\ldots,p$, need to be described to fully specify the multivariate
random field.
Authors typically refer to $C_{ij}$ as direct- or marginal-covariance
functions for
$i=j$, and cross-covariance functions for $i\neq j$.
Here, we assume that $\bZ(\bs)$ is a mean zero process.
The quantities $\rho_{ij}(\bs_1,\bs_2)=C_{ij}(\bs_1,\bs_2)/\{
C_{ii}(\bs_1,\bs_1)\cdot\break C_{jj}(\bs_2,\bs_2)\}^{1/2}$
are the cross-correlation functions.
Our goal is then to construct valid and flexible cross-covariance
functions (\ref{ccdef}),
that is, the matrix-valued mapping $\bC\dvtx  \R^d \times\R^d
\rightarrow M_{p \times p}$,
where $M_{p \times p}$ is the set of $p \times p$ real-valued matrices,
must be nonnegative definite in the following sense. The covariance
matrix $\bSigma$ of the random vector $\{\bZ(\bs_1)^{\mathrm
{T}},\ldots,\bZ
(\bs_n)^{\mathrm{T}}\}^{\mathrm{T}}\in\R^{np}$:
%
%
%e2 #&#
\begin{equation}
\label{bSigma}\hspace*{13pt} \bSigma= %
\pmatrix{ \bC(\bs_1,
\bs_1)& \bC(\bs_1,\bs_2) & \cdots& \bC(
\bs_1,\bs_n)
\cr
\bC(\bs_2,
\bs_1) & \bC(\bs_2,\bs_2) & \cdots& \bC(
\bs_2,\bs_n)
\cr
\vdots& \vdots& \ddots& \vdots
\cr
\bC(
\bs_n,\bs_1)& \bC(\bs_n,\bs_2) &
\cdots& \bC(\bs_n,\bs_n) } %
,
\end{equation}
should be nonnegative definite: $\mathbf{a}^{\mathrm{T}}\bSigma
\mathbf{a}\geq
0$ for any
vector $\mathbf{a}\in\R^{np}$, any spatial locations $\bs_1,\ldots
,\bs_n$, and
any integer $n$. \citet{FanDig12} reviewed approaches for the
bivariate
case $p=2$, although most techniques can be readily extended to $p>2$, and
\citet{AlvRosLaw12} reviewed approaches for machine learning.

A multivariate random field is second-order stationary (or just
stationary) if
the marginal and cross-covariance
functions depend only on the separation vector $\bh=\bs_1-\bs_2$,
that is,
there is a mapping $C_{ij}\dvtx\allowbreak   \R^d \rightarrow\R$ such that
\[
\cov \bigl\{Z_i(\bs_1),Z_j(
\bs_2) \bigr\}=C_{ij}(\bh), \quad \bh\in\R^d.
\]
Otherwise, the process is nonstationary.
Stationarity can be thought of as an invariance property under the
translation of
coordinates. A test for the stationarity of a multivariate random field
can be
found in \citet{JunGen12}.

A multivariate random field is isotropic if it is stationary and
invariant under
rotations and reflections, that is,
there is a mapping $C_{ij}\dvtx  \R_+\cup\{0\} \rightarrow\R$ such that
\[
\cov \bigl\{Z_i(\bs_1),Z_j(
\bs_2) \bigr\}=C_{ij}\bigl(\|\bh\|\bigr), \quad \bh\in\R^d,
\]
where $\|\cdot\|$ denotes the Euclidean norm. Otherwise, the
multivariate random
field is anisotropic. Isotropy or even stationarity are not always realistic,
especially for large spatial regions,
but sometimes are satisfactory working assumptions and serve as basic elements
of more sophisticated anisotropic and nonstationary models.

In the univariate setting, variograms are often the main focus in geostatistics,
and are defined as the variance of contrasts.
Variograms can be extended to multivariate random fields in two ways:
A covariance-based cross-variogram \citep{Mye82} defined as
%
%
%e3 #&#
\begin{eqnarray}
\label{crossvario} \cov \bigl\{Z_i(\bs_1)-Z_i(
\bs_2),Z_j(\bs_1)-Z_j(
\bs_2) \bigr\},
\end{eqnarray}
$\bs_1,\bs_2 \in
\R^d$, and a variance-based cross-variogram \citep{Mye91}, also coined pseudo
cross-variogram,
%
%
%e4 #&#
\begin{equation}
\label{pseudocrossvario} \var \bigl\{Z_i(\bs_1)-Z_j(
\bs_2) \bigr\},\quad  \bs_1,\bs_2 \in
\R^d.
\end{equation}
The corresponding stationary versions are immediate. \citet{CreWik98} reviewed the differences
between (\ref{crossvario}) and (\ref{pseudocrossvario}), and argued
that (\ref{pseudocrossvario}) is more appropriate for co-kriging
because it yields the same optimal co-kriging predictor as the one
obtained with the cross-covariance function $C_{ij}$ in
(\ref{ccdef}); see also \citet{VerCre93} and \citet{Huaetal09}.
Unfortunately, the interpretation of cross-variograms is difficult, and
so most
authors favor working with covariance and cross-covariance formulations.

%%%%%%%%%%%%%%%%%%%%%%%%%%%%%%%%%%%%%%%%%%%%%%%%%%%%%%%%%%%%%%%%%%%%%%%%
%s1.2 #&#
\subsection{Properties of Cross-Covariance Matrix Functions}\label
{sec:crosscov}

Because the covariance matrix $\bSigma$ in (\ref{bSigma}) must be
symmetric, the
matrix functions must
satisfy $\bC(\bs_1,\bs_2)=\bC(\bs_2,\bs_1)^{\mathrm{T}}$, or
$\bC(\bh
)=\bC(-\bh)^{\mathrm{T}}$ under
stationarity.\linebreak[4]  Therefore, cross-covariance matrix functions are not
symmetric in
general, that is,
\begin{eqnarray*}
C_{ij}(\bs_1,\bs_2)&=&\cov \bigl
\{Z_i(\bs_1),Z_j(\bs_2) \bigr\}
\\
&\neq&\cov \bigl\{ Z_j(\bs_1),Z_i(
\bs_2) \bigr\}=C_{ji}(\bs_1,\bs_2),
\end{eqnarray*}
$\bs_1,\bs_2 \in\R^d$, unless the cross-covariance functions themselves are all symmetric
(\citeauthor{kse03}, \citeyear{kse03}).
However, the collocated matrices $\bC(\bs,\bs)$, or $\bC(\0)$ under
stationarity, are symmetric and nonnegative definite.

The marginal and cross-covariance functions satisfy
$|C_{ij}(\bs_1,\bs_2)|^2\leq C_{ii}(\bs_1,\bs_1) C_{jj}(\bs_2,\bs
_2)$, or\break
$|C_{ij}(\bh)|^2\leq C_{ii}(\0) C_{jj}(\0)$ under stationarity.
However, $|C_{ij}(\bs_1,\bs_2)|$ need not be less than or equal
to $C_{ij}(\bs_1,\bs_1)$, or $|C_{ij}(\bh)|$ need not be less than
or equal to
$C_{ij}(\0)$ under stationarity. This is because
the maximum value of $C_{ij}(\bh)$ is not restricted to occur at $\bh
=\0$,
unless $i=j$, and in fact this sometimes occurs in practice \citep{LiZha11}.
Thus, there are no similar bounds between $|C_{ij}(\bs_1,\bs_2)|^2$ and
$C_{ii}(\bs_1,\bs_2) C_{jj}(\bs_1,\bs_2)$, or between
$|C_{ij}(\bh)|^2$ and $C_{ii}(\bh) C_{jj}(\bh)$ under stationarity.

A cross-covariance matrix function is separable if
%
%
%e5 #&#
\begin{equation}
\label{ccSep} C_{ij}(\bs_1,\bs_2)=\rho(
\bs_1,\bs_2) R_{ij},\quad  \bs_1,
\bs_2 \in\R^d, %\cov(Z_i,Z_j)
\end{equation}
for all $i,j=1,\ldots,p$, where $\rho(\bs_1,\bs_2)$ is a valid,
nonstationary or
stationary, correlation function and $R_{ij} = \cov(Z_i,Z_j)$ is the nonspatial
covariance between variables $i$ and $j$. \citet{MarGoo93} introduced
and used separability to model multivariate\linebreak[4]  spatio-temporal data, and
\citet{autokey9} used separable covariances in the context of
computer model
calibration. In the past, separable cross-covariance structures were
sometimes called intrinsic coregionalizations \citep{HelCre94}.

With a large number of processes, detecting structures of the multivariate
random process such as symmetry and separability can be difficult via
elementary data analytic techniques.
\citet{LiGenShe08} proposed an approach based on the asymptotic
distribution of the
sample cross-covariance estimator to test these various structures.
Their methodology allows the practitioner to assess the underlying dependence
structure of the data and to suggest appropriate cross-covariance
functions, an
important part of model building.

In the special case of stationary matrix-valued covariance functions, there
is an intimate link between the cross-covariance matrix function and its
spectral representation. In particular, define the cross-spectral densities
$f_{ij}\dvtx \R^d \to\R$ as
\[
f_{ij}(\bomega) = \frac{1}{(2\pi)^d}\int_{\R^d}
e^{-\iota\bh^{\mathrm{T}}
\bomega} C_{ij}(\bh) \,\mathrm{d}\bh, \quad \bomega\in\R^d,
\]
where $\iota=\sqrt{-1}$ is the imaginary number.
A necessary and sufficient condition for $\bC(\cdot)$ to be a
valid (i.e., nonnegative definite), stationary matrix-valued covariance function
is for the matrix function $\bff(\bomega_0)= \{f_{ij}(\bomega_0)\}
_{i,j=1}^p$ to be nonnegative\vadjust{\goodbreak}
definite for any $\bomega_0$ \citep{Cra40}. While Cram\'er's
original result is stated in terms of
measures of bounded variation, in practice using spectral densities is
preferred.
This can be viewed as a multivariate extension of Bochner's celebrated
theorem \citep{Boc55}.
The analogue of Schoenberg's theorem for multivariate random fields,
that is, Bochner's theorem for isotropic cross-covari\-ance functions,
has recently been
investigated by \citeauthor{AloPorGir} (\citeyear{AloPorGir}, \citeyear{autokey3}).

%%%%%%%%%%%%%%%%%%%%%%%%%%%%%%%%%%%%%%%%%%%%%%%%%%%%%%%%%%%%%%%%%%%%%%%%
%s1.3 #&#
\subsection{Estimation of Cross-Covariances}

The empirical estimator of the cross-covariance matrix function of a
stationary multivariate random field is
%
%
%e6 #&#
\begin{eqnarray}\label{empcrosscovmat}
\wh\bC(\bh)&=&\frac{1}{|N(\bh)|} \sum_{(k,l)\in N(\bh)} \bigl\{
\bZ( \bs_k)-\bar\bZ \bigr\}
\nonumber
\\[-8pt]
\\[-8pt]
\nonumber
&&\hspace*{72pt}{}\cdot\bigl\{\bZ(\bs_l)-\bar\bZ
\bigr\} ^{\mathrm{T}},
\end{eqnarray}
$\bh\in\R^d$, where $N(\bh)=\{(k,l)| \bs_k-\bs_l=\bh\}$, $|N(\bh)|$\vspace*{1.5pt} denotes its
cardinality, and $\bar\bZ=\frac{1}{n}\sum_{k=1}^n \bZ(\bs_k)$ is
the sample mean vector.
A valid parametric model is then typically fit by least squares methods
to the empirical estimates in (\ref{empcrosscovmat}). Alternatively,
one can use likelihood-based methods or the Bayesian paradigm \citep{BroLeZid94}.
In any case, valid and flexible cross-covariance models are needed.
\citet{KunPapBas97} studied generalized cross-covariances and their
estimation.

\citet{PapKunWeb93} discussed empirical estimators of the
cross-variogram (\ref{crossvario}) and (\ref{pseudocrossvario}).
Unlike the pseudo cross-variogram, the cross-\linebreak[4] variogram (\ref
{crossvario}) has the disadvantage that it cannot be estimated
when the variables are not observed at the same spatial locations.
\citet{Lar03} proposed two outlier-robust estimators of the pseudo
cross-variogram (\ref{pseudocrossvario}) and applied them
in a multivariate geostatistical analysis of soil properties.
\citet{Fur05} studied the bias of the empirical cross-covariance matrix
$\bC(\mathbf{0})$ estimation under spatial dependence using
both fixed-domain and increasing-domain asymptotics.
\citet{LimSte08} investigated a spectral approach based on spatial
cross-periodograms for data on a lattice and studied their properties
using fixed-domain asymptotics.

%%%%%%%%%%%%%%%%%%%%%%%%%%%%%%%%%%%%%%%%%%%%%%%%%%%%%%%%%%%%%%%%%%%%%%%%
%s2 #&#
\section{Cross-Covariances Built from Univariate Models}\label{sec:LMC}
%%%%%%%%%%%%%%%%%%%%%%%%%%%%%%%%%%%%%%%%%%%%%%%%%%%%%%%%%%%%%%%%%%%%%%%%

The most common approach to building cross-covariance functions is by
combining univariate covariance functions. The three main options in this
vein are the linear model of coregionalization, various convolution techniques
and the use of latent dimensions.\vadjust{\goodbreak}

%%%%%%%%%%%%%%%%%%%%%%%%%%%%%%%%%%%%%%%%%%%%%%%%%%%%%%%%%%%%%%%%%%%%%%%%
%s2.1 #&#
\subsection{Linear Model of Coregionalization}

Probably the most popular approach of combining univariate covariances is
the so-called linear model of coregionalization
(LMC) for stationary random fields (\cite*{BouMar91};
\citeauthor{GouVol92}, \citeyear{GouVol92}; \cite*{GrzWac94};
\cite*{VarWarMye02};
\cite*{SchGel}; \citeauthor{kse03}, \citeyear{kse03}).
It consists of representing the multivariate
random field as a linear combination of $r$ independent univariate
random fields.
The resulting cross-covariance functions take the form
%
%
%e7 #&#
\begin{equation}\label{ccLMC}
C_{ij} (\bh)=\sum_{k=1}^r
\rho_k(\bh)A_{ik}A_{jk},\quad  \bh\in\R^d,
\end{equation}
for an integer $1\leq r \leq p$, where $\rho_k(\cdot)$ are valid stationary
correlation functions and $\bA= (A_{ij})^{p,r}_{i,j=1}$
is a $p \times r$ full rank matrix. When $r=1$, the cross-covariance function
(\ref{ccLMC}) is separable as in (\ref{ccSep}). The allure of this approach
is that only $r$ univariate covariances $\rho_k(\bh)$ must be specified,
thus avoiding direct specification of a valid cross-covariance matrix function.
The LMC can additionally be built from a conditional perspective
(\cite{RoyBer99}; \cite{Geletal04}). Note that the discrete
sum representation (\ref{ccLMC}) can also be interpreted as a scale
mixture \citep{PorZas11}.

With a large number of processes, the number of parameters can quickly become
unwieldy and the resulting estimation difficult.
\citet{Zha07} described maximum likelihood estimation of the spatial
LMC based on an EM algorithm,
whereas \citet{SchGel} proposed a Bayesian coregionalization
approach with application to multivariate pollutant data.
A second drawback of the LMC is that the smoothness of any component
of the multivariate random field is restricted to that of the roughest
underlying
univariate process.

%%%%%%%%%%%%%%%%%%%%%%%%%%%%%%%%%%%%%%%%%%%%%%%%%%%%%%%%%%%%%%%%%%%%%%%%
%s2.2 #&#
\subsection{Convolution Methods}

Convolution methods fall into the two categories of kernel and
covariance convolution.
The kernel convolution method (\cite*{VerBar98};  \cite*{VerCreBar04}) uses
\begin{eqnarray}
&&C_{ij}(\bh)\nonumber\\
&&\quad =\int_{\R^d}\int_{\R^d}k_i(
\bv_1)k_j(\bv_2)\rho(\bv_1-
\bv_2+\bh)\,\mathrm{d}\bv_1\,\mathrm{d}\bv_2,\nonumber
\end{eqnarray}
$\bs_1,\bs_2 \in\R^d$, where the $k_i$ are square integrable kernel functions and $\rho(\cdot
)$ is a
valid stationary correlation function.
This approach assumes that all the spatial processes $Z_i(\bs)$, for
$i=1,\ldots,p$,
are generated by the same underlying process, which is very restrictive
in that it
imposes strong dependence between all constituent processes $Z_i(\bs
)$. Overall,
this approach and its parameters can be difficult to interpret and,
except for some special cases, requires numerical integration.

Covariance convolution for stationary spatial random fields
(\cite{GasCoh99}; \cite{Gasetal}; \cite{MajGel07}) yields
\[
C_{ij}(\bh)=\int_{\R^d} C_i(\bh-
\bk)C_j(\bk)\,\mathrm{d}
\bk, \quad \bh\in\R^d,
\]
where $C_i$ are square integrable functions.
Although some closed-form expressions exist, this method usually
requires numerical integration. A particularly useful example of a closed form
solution is
when the $C_i$ are Mat\'ern correlation functions with common scale parameters.
In this setup, Mat\'ern correlations are
closed under convolution and this approach results in a special
case of the multivariate Mat\'ern model \citep{GneKleSch10}.

%%%%%%%%%%%%%%%%%%%%%%%%%%%%%%%%%%%%%%%%%%%%%%%%%%%%%%%%%%%%%%%%%%%%%%%%
%s2.3 #&#
\subsection{Latent Dimensions}\label{sec:latdim}

Another approach to build valid cross-covariance functions based on univariate
($p=1$) spatial covariances was put forward by \citet{ApaGen10}
(see also \cite*{PorZas11}).
Their idea was to create additional latent dimensions that represent
the various variables to be modeled. Specifically, each component $i$
of the multivariate random field $\bZ(\bs)$ is represented
as a point $\bxi_i=(\xi_{i1},\ldots,\xi_{ik})^{\mathrm{T}}$ in $\R^k$,
$i=1,\ldots,p$, for an integer $1\leq k \leq p$, yielding
the marginal and cross-covariance functions
%
%
%e8 #&#
\begin{eqnarray}\label{latgen}
\hspace*{20pt}C_{ij}(\bs_1,\bs_2)=C \bigl\{(
\bs_1,\bxi_i),(\bs_2,\bxi_j)
\bigr\}, \quad \bs_1,\bs_2 \in\R^d,
\end{eqnarray}
where $C$ is a valid univariate covariance function on $\R^{d+k}$; see
\citet{autokey46} for a review of possible univariate
covariance functions. It is immediate that the resulting
cross-covariance matrix
$\bSigma$ in (\ref{bSigma}) is nonnegative definite because its
entries are defined through a valid univariate covariance.
If the covariance $C$ is from a stationary or isotropic univariate
random field, then so is also the cross-covariance
function (\ref{latgen}); for instance, $C_{ij}(\bh)=C(\bh,\bxi
_i-\bxi_j)$.

As an example of the aforementioned construction, \citet{ApaGen10}
suggested
%
%
%e9 #&#
\begin{eqnarray}\label
{latgenexample}
\hspace*{18pt}C_{ij}(\bh)&=&\frac{\sigma_i \sigma_j}{\|\bxi_i-\bxi_j\|+1}\exp \biggl\{ \frac{-\alpha\|\bh\|}{(\|\bxi_i-\bxi_j\|+1)^{\beta
/2}} \biggr\}\nonumber\\[-8pt]\\[-8pt]
&&{}+\tau^2 I(i=j)I(\bh=\mathbf{0}), \quad \bh\in\R^d,
\nonumber
\end{eqnarray}
where $I(\cdot)$ is the indicator function, $\sigma_i>0$ are margi\-nal
standard deviations, $\tau\geq0$ is a nugget effect, and $\alpha>0$
is a length scale.
Here, $\beta\in[0,1]$ controls the nonseparability between space and
variables, with $\beta=0$ being the separable case.
The parameters of the model are estimated by maximum likelihood or
composite likelihood methods.
\citet{ApaGen10} provided an application to a trivariate
pollution dataset from California.
Further use of latent dimensions for multivariate spatio-temporal random
fields are discussed in Section~\ref{sec:spatiotemp}. The idea of
latent dimensions was
recently extended to modeling nonstationary processes by \citet{BorShaZid12}.

%%%%%%%%%%%%%%%%%%%%%%%%%%%%%%%%%%%%%%%%%%%%%%%%%%%%%%%%%%%%%%%%%%%%%%%%
%s3 #&#
\section{Mat\'ern Cross-Covariance Functions}
%%%%%%%%%%%%%%%%%%%%%%%%%%%%%%%%%%%%%%%%%%%%%%%%%%%%%%%%%%%%%%%%%%%%%%%%

The Mat\'ern class of positive definite functions has become the standard
covariance model for univariate fields (\citeauthor{GutGne06}, \citeyear{GutGne06}).
The popularity in large part is
due to the work of \citet{Ste99} who showed that the behavior of the covariance
function near the origin has fundamental implications on predictive
distributions,
particularly predictive uncertainty. The key feature of the Mat\'ern is the
inclusion of a smoothness parameter that directly controls correlation at
small distances. The Mat\'ern correlation function is
\begin{eqnarray*}
\mathrm{M}(\bh| \nu, a) = \frac{2^{1-\nu}}{\Gamma(\nu)} \bigl(a\|\bh \|\bigr)^\nu
\mathrm{K}_\nu\bigl(a\|\bh\|\bigr), \quad \bh\in\R^d,
\end{eqnarray*}
where $\mathrm{K}_\nu$ is a modified Bessel function of order $\nu$,
$a>0$ is a length scale
parameter that controls the rate of decay of
correlation at larger distances, while $\nu>0$ is the smoothness
parameter that
controls behavior of correlation near the origin.
The smoothness parameter is aptly named as it implies levels of mean square
differentiability of the random process, with large $\nu$ yielding very
smooth processes that are many times differentiable, and small $\nu$ yielding
rough processes; in fact there is a direct connection between the smoothness
parameter and the Hausdorff dimension of the resulting random process
\citep{GofJor88}.

Due to its popularity for univariate modeling, there is interest in being
able to simultaneously model multiple processes, each of which marginally
has a Mat\'ern correlation structure. To this end, \citet{GneKleSch10}
introduced the so-called multivariate Mat\'ern model, where each constituent
process is allowed a marginal Mat\'ern correlation, with Mat\'erns also
composing the cross-correlation structures. In particular, the multivariate
Mat\'ern implies
%
%
%e10 #&#
\begin{eqnarray}
\label{eq:mm} \rho_{ii}(\bh) &=& \mathrm{M}(\bh| \nu_i,a_i)
\quad \mbox{and}\nonumber\\[-8pt]\\[-8pt]
 \rho_{ij}(\bh) &=& \beta_{ij}\mathrm{M}(\bh|
\nu_{ij},a_{ij}), \quad \bh\in\R^d.\nonumber
\end{eqnarray}
Of course, this correlation structure can be coerced to a covariance structure
by multiplying $C_{ii}(\bh)$ by $\sigma_i^2$ and $C_{ij}(\bh)$ by
$\sigma_i \sigma_j$. Here, $\beta_{ij}$ is a collocated cross-correlation
coefficient, and represents the strength of correlation between $Z_i$ and
$Z_j$ at the same location, $\bh= \mathbf{0}$.

The difficulty in (\ref{eq:mm}) is deriving conditions on model parameters
$\nu_i,\nu_{ij},a_i,a_{ij}$ and $\beta_{ij}$ that result in a valid,
that is, a
nonnegative definite multivariate covariance class. In the original work,
\citet{GneKleSch10} described two main models, the \emph{parsimonious
Mat\'ern} and the \emph{full bivariate Mat\'ern}. The parsimonious
Mat\'ern
is a reduction in complexity over (\ref{eq:mm}) in that $a_i = a_{ij}
= a$ are
held at the same value for all marginal and cross-covariances, and the
cross-smoothnesses are set to the arithmetic average of the marginals,
$\nu_{ij} = (\nu_i + \nu_j)/2$. The model is then valid with an easy-to-check
condition on the cross-correlation coefficient $\beta_{ij}$.

The flexibility of the parsimonious Mat\'ern is in allowing each process
to have a distinct marginal smoothness behavior, and thus allowing for
simultaneous modeling of highly smooth and rough fields. The natural extension
to allow distinct process-dependent length scale parameters $a_i$ turns
out to
be more involved. The full bivariate Mat\'ern of \citet{GneKleSch10}
allows for distinct smoothness and scale parameters for two processes
(and in
fact results in a characterization for $p=2$). A second set of authors,
\citet{ApaGenSun12}, were able to overcome the deficiencies of the
parsimonious formulation for $p>2$, introducing the \emph{flexible
Mat\'ern}.
The flexible Mat\'ern works for any number of processes, allowing for
each process to have distinct smoothness and scale parameters, and is
as close in spirit to allowing entirely free marginal Mat\'ern covariances
with some level of cross-process dependence as is currently available.
A number of simpler sufficient conditions are available by using scale
mixtures (\cite*{ReiBur07};
\cite*{GneKleSch10}; \cite*{Sch10}; \cite*{PorZas11}).

It is worth pointing out that the experimental results of both sets of
authors, \citet{GneKleSch10} and \citet{ApaGenSun12},
highlighted the
importance of allowing for highly flexible and distinct marginal covariance
structures, while still allowing for some degree of cross-process correlation,
and indeed the improvement over an independence assumption was substantial.

%%%%%%%%%%%%%%%%%%%%%%%%%%%%%%%%%%%%%%%%%%%%%%%%%%%%%%%%%%%%%%%%%%%%%%%%
%s4 #&#
\section{Nonstationary Cross-Covariance Functions}
%%%%%%%%%%%%%%%%%%%%%%%%%%%%%%%%%%%%%%%%%%%%%%%%%%%%%%%%%%%%%%%%%%%%%%%%

Geophysical, environmental and ecological spatial processes often
exhibit spatial
dependence that depends on fixed geographical features such as terrain or
land use type, or dynamical environments such as prevailing winds.
In either case, the evolving nature of spatial dependence
is not well captured by stationary models, and thus the availability of
nonstationary constructions is desired, that is, models such that the
marginal and cross-covariance functions are now dependent on the spatial
location pair, not just the lag vector,
$\cov\{Z_i(\bs_1),Z_j(\bs_2)\} = C_{ij}(\bs_1,\bs_2)$.

Many of the aforementioned models have been extended to the nonstationary
setup, including the original stationary models as special cases.
The first natural extension to allowing the LMC to be nonstationary
is to let the latent univariate correlations be nonstationary, so that
\begin{eqnarray*}
C_{ij}(\bs_1,\bs_2) = \sum
_{k=1}^r \rho_{k}(\bs_1,
\bs_2) A_{ik}A_{jk}, \quad \bs_1,
\bs_2 \in\R^d,
\end{eqnarray*}
where now $\rho_{k}$ are nonstationary univariate correlation functions.
The onus of deriving a matrix-valued nonstationary covariance function
is then
alleviated in favor of opting for univariate nonstationary correlations,
of which there are many choices (e.g., \cite*{SamGut92}; \cite*{Fue02}; \cite*{PacSch06}; \cite*{BorShaZid12}).
Although this extension seems straightforward,
we are unaware of any authors who have implemented such an approach.
The second way to extend the LMC to a nonstationary setup is to allow the
coefficients to be spatially varying \citep{Geletal04}, so that
\begin{eqnarray}
C_{ij}(\bs_1,\bs_2) = \sum
_{k=1}^r \rho_{k}(\bs_1-
\bs_2) A_{ik}(\bs_1)A_{jk}(
\bs_2), \nonumber
\end{eqnarray}
$\bs_1,\bs_2 \in\R^d$. This type of approach can be useful if the observed multivariate process
is linked in a varying way to some underlying and unobserved processes.
\citet{Guhetal13} combined a low rank predictive process
approach with
the nonstationary LMC for computationally feasible modeling with large datasets.

The multivariate Mat\'ern was extended to the nonstationary case by
\citet{KleNyc12}. The basic idea is to allow the various Mat\'ern
parameters, variance, smoothness and length scale, to be spatially varying
(\cite*{Ste05}; \cite*{PacSch06}), using normal scale mixtures
\citep{Sch10}. For example, temperature fields
exhibit longer range spatial dependence over the ocean than over land
due to
terrain driven nonstationarity, and a nonstationary Mat\'ern\vadjust{\goodbreak} with spatially
varying length scale parameter can capture this type of dependence without
resorting to using disjoint models between ocean and land. In particular,
the nonstationary multivariate Mat\'ern supposes
\begin{eqnarray}
\rho_{ii}(\bs_1,\bs_2)&\propto&\mathrm{M}
\bigl(\bs_1,\bs_2 | \nu_i(
\bs_1,\bs_2),a_i(\bs_1,
\bs_2) \bigr),
\nonumber\\
\rho_{ij}(\bs_1,\bs_2) &\propto&
\beta_{ij}(\bs_1,\bs_2)\mathrm{M} \bigl(
\bs_1,\bs_2 | \nu_{ij}(\bs_1,
\bs_2),a_{ij}(\bs_1,\bs_2) \bigr),\nonumber
\end{eqnarray}
$\bs_1,\bs_2 \in\R^d$. An additional point here is that $\beta_{ij}(\bs,\bs)$ is
proportional to the
collocated cross-correlation coefficient $\mathrm{cor}\{Z_i(\bs
),Z_j(\bs)\}$, that is, the
strength of relationship between variables at the same location. This strength
often varies spatially, for example minimum and maximum temperature are less
correlated over highly mountainous regions than over plains where they exhibit
greater dependence. \citet{KleGen13} considered an approach to allowing
this correlation coefficient to vary with location in such a way that
it can be
included with any arbitrary multivariate covariance
choice, as long as each process has a nonzero nugget effect (which is not
usually restrictive, as most processes exhibit small scale dependence that
are typically modeled as nugget effects). Other authors have noted similar
phenomena with other scientific data (\cite*{FueRei13}; \cite*{Guhetal13}).

Owing to the increasing complexity of nonstationary and multivariate models
and the expertise required to decide on a framework as well as
implement an
estimation scheme, a few authors have considered nonparametric approaches
to estimation. Extending \citet{Oeh93} and \citet{GuiSenMon01} to the
multivariate case,
\citet{Junetal11} and \citet{KleKatRaj13} worked with a nonparametric
estimator
of multivariate covariance that is free from model choice and is available
throughout the observation domain. The underlying idea is to kernel smooth
the empirical method-of-moments estimate of spatial covariance in a way
that retains nonnegative definiteness and yields covariance estimates
at any
arbitrary location pairs, not only those with observations. Their nonparametric
estimators are variations on the form
%
%
%e11 #&#
\begin{eqnarray}
\label{eq:kern} &&\hat{C}_{ij}(\bx,\by) \nonumber\\
&&\quad = \Biggl(
\sum_{k=1}^n \sum_{\ell=1}^n K_\lambda\bigl(\|\bx- \bs_k\|\bigr)\nonumber\\
&&\hphantom{\quad = \Biggl(
\sum_{k=1}^n \sum_{\ell=1}^n}{}\cdot
K_\lambda\bigl(\|\by- \bs_\ell\|\bigr) Z_i(\bs_k) Z_j(\bs_\ell)\Biggr)\\
&&\qquad {}\cdot\Biggl(
\sum_{k=1}^n \sum_{\ell=1}^n K_\lambda\bigl(\|\bx- \bs_k\|\bigr)
K_\lambda\bigl(\|\by- \bs_\ell\|\bigr)\Biggr)^{-1}, \nonumber
\end{eqnarray}
$\bx,\by\in\R^d$, where $K_\lambda(r) = K(r/\lambda)$ is a positive kernel function
with bandwidth
$\lambda$. The displayed equation (\ref{eq:kern}) is set up for the
case when
$Z_i$ is mean zero for $i=1,\ldots,p$, for instance representing residuals
after a mean trend has been removed; the estimator can also be applied to
centered residuals such as $Z_i(\bs_k) - \bar{Z}_i$.
This type of estimator can capture substantial
nonstationarity that may be difficult to pick up parametrically
\citep{KleKatRaj13}. The nonparametric approach to estimation is primarily
useful when
replications of the multivariate random field are available. Although
it can
be applied when only a single field realization is available, we
caution against
its use given the well-known variability of empirical estimates in
small samples.

The two methods of covariance and kernel convolution can also
be extended to result in nonstationary matrix functions (\citeauthor{Cal07}, \citeyear{Cal07}, \citeyear{Cal08};
\cite*{MajPauBau10}). As with the univariate case, the convolution integrals
are often intractable and must be estimated numerically, and parametric
interpretations are sometimes ambiguous.

%%%%%%%%%%%%%%%%%%%%%%%%%%%%%%%%%%%%%%%%%%%%%%%%%%%%%%%%%%%%%%%%%%%%%%%%
%s5 #&#
\section{Cross-Covariance Functions with Special Features}
%%%%%%%%%%%%%%%%%%%%%%%%%%%%%%%%%%%%%%%%%%%%%%%%%%%%%%%%%%%%%%%%%%%%%%%%

%%%%%%%%%%%%%%%%%%%%%%%%%%%%%%%%%%%%%%%%%%%%%%%%%%%%%%%%%%%%%%%%%%%%%%%%
%s5.1 #&#
\subsection{Asymmetric Cross-Covariance Functions}

All the stationary models described so far are symmetric, in the sense
that $C_{ij}(\bh)=C_{ji}(\bh)$,
or equivalently, $C_{ij}(\bh)=C_{ij}(-\bh)$. Although $C_{ij}(\bh
)=C_{ji}(-\bh)$ by
definition, the aforementioned
properties may not hold in general. \citet{LiGenShe08} proposed a test of
symmetry of the cross-covariance
structure of multivariate random fields based on the asymptotic distribution
of its empirical estimator. If the test
rejects symmetry, then asymmetric cross-covariance functions are needed.

\citet{LiZha11} proposed a general approach to render any
stationary symmetric
cross-covariance function asymmetric.
The key idea is to notice that if $C_{ij}(\bh)$ is a valid symmetric
cross-covariance function, then
%
%
%e12 #&#
\begin{eqnarray}\label{asymcross}
C_{ij}^a(\bh)=C_{ij}(\bh+\mathbf{a}_i-
\mathbf{a}_j),\quad  \bh\in\R^d,
\end{eqnarray}
is a valid asymmetric cross-covariance function for any vectors
$\mathbf{a}_i \in\R^d$, $i=1,\ldots,p$, such that
$\mathbf{a}_i\neq\mathbf{a}_j$. Indeed, if $\bZ(\bs)=\{Z_1(\bs
),\ldots,Z_p(\bs)\}^{\mathrm{T}}$
has cross-covariance functions $C_{ij}(\bh)$, then
$\{Z_1(\bs-\mathbf{a}_1),\ldots,\allowbreak Z_p(\bs-\mathbf{a}_p)\}^{\mathrm
{T}}$ has
cross-covariance functions
$C_{ij}^a(\bh)$ given by (\ref{asymcross}),
$i,j=1,\ldots,p$. In particular, the construction (\ref{asymcross})
can be
used to produce asymmetric versions of the LMC and the multivariate
Mat\'ern models. The vectors $\mathbf{a}_1,\ldots,\mathbf{a}_p$
introduce delays that
generate asymmetry in the cross-covariance
structure. Because only the differences $\mathbf{a}_i-\mathbf{a}_j$
matter, one can impose
a constraint such as $\mathbf{a}_1+\cdots+\mathbf{a}_p=\mathbf{0}$
or $\mathbf{a}_1=\mathbf{0}$ to ensure
identifiability. \citet{LiZha11} proposed to first estimate the
marginal parameters of $C_{ij}^a(\bh)$ in (\ref{asymcross}), and then estimate
the cross-parameters and $p-1$ of the $\mathbf{a}_i$'s. Their
simulations and data
examples showed that asymmetric cross-covariance functions, when
required, can achieve
remarkable improvements in prediction over symmetric models.
\citet{ApaGen10} used a similar strategy to produce asymmetric
spatio-temporal cross-covariance models based on latent dimensions; see
Section~\ref{sec:spatiotemp}. Inducing asymmetry in a nonstationary
model is
yet an
open problem.

%%%%%%%%%%%%%%%%%%%%%%%%%%%%%%%%%%%%%%%%%%%%%%%%%%%%%%%%%%%%%%%%%%%%%%%%
%s5.2 #&#
\subsection{Compactly Supported Cross-Covariance Functions}

Computational issues in the face of large datasets is a major problem
in any spatial analysis, including likelihood calculations and/or co-kriging;
see the review by \citeauthor{SunLiGen} (\citeyear{SunLiGen}, Section~3.7). Especially, if the observation
network is very large (even on the order of thousands), likelihood calculations
and co-kriging equations are difficult or impossible to solve with standard
covariance models, due to the dense unstructured observation covariance
matrix. One approach to overcoming this difficulty is to induce
sparsity in
the covariance matrix, either by using a compactly supported covariance function
as the model, or by covariance tapering, that is, multiplying a
compactly supported
nonnegative definite function against the model covariance (\cite*{FurGenNyc06};
\cite*{KauSchNyc08}). Then sparse matrix methods can be used to invert the
covariance matrix, or find the determinant thereof.

Only recently have authors begun to consider this problem for multivariate
random fields. Most
of the currently available models are based on scale mixtures of the form
\[
C_{ij}(\bh) = \int\bigl(1-\|\bh\|/x\bigr)^{\nu}_+ g_{ij}(x)
\,\mathrm{d}x,\quad  \bh\in\R^d,
\]
or variations on this theme (\cite*{ReiBur07}; \cite*{PorZas11}).
Here,\vspace*{2pt} $\nu\geq(d+1)/2$, and
$\{g_{ij}(x)\}_{i,j=1}^p$ forms a valid cross-covariance matrix function.
The generality of this construction gives rise to many interesting examples.
For instance, with $g_{ij}(x) = x^{\nu} (1-x/b)_+^{\gamma_{ij}}$ where
$\gamma_{ij} = (\gamma_i + \gamma_j)/2$ and $\gamma_i>0$ for all
$i=1,\ldots,p$
we have the multivariate Askey taper
\begin{eqnarray*}
C_{ij}(\bh) &=& b^{\nu+ 1} B(\gamma_{ij}+1,\nu+1)\biggl(1-\frac{\|\bh\|}{b} \biggr)^
{\nu+\gamma_{ij} + 1},
\end{eqnarray*}
$\|\bh \| < b$, and $0$ otherwise, where $B$ is the beta function \citep{Poretal13}.
\citet{KlePor} provided a nonstationary extension of this model,
while \citet{Poretal13} considered similar ideas for Buhmann
functions and
B-splines. \citet{DalPorBev} obtained multivariate Askey functions
with different compact supports $b_{ij}$ and the multivariate analogue of
Wendland functions.
The latter provide a tool for tapering cross-covariance functions such
as the multivariate Mat\'ern.
Recent results on equivalence of Gaussian measures of multivariate
random fields by
\citet{RuiPor} will allow for assessing the statistical
properties of multivariate tapers.
\citet{DuMa13} derived compactly supported classes
of the P\'olya type. Although there has been a flurry of recent activity,
much additional work remains in implementing these models in real world
applications, exploring covariance
tapering and understanding limitations of stationary constructions.

%
%f1 #&#
\begin{figure*}[b]

\includegraphics{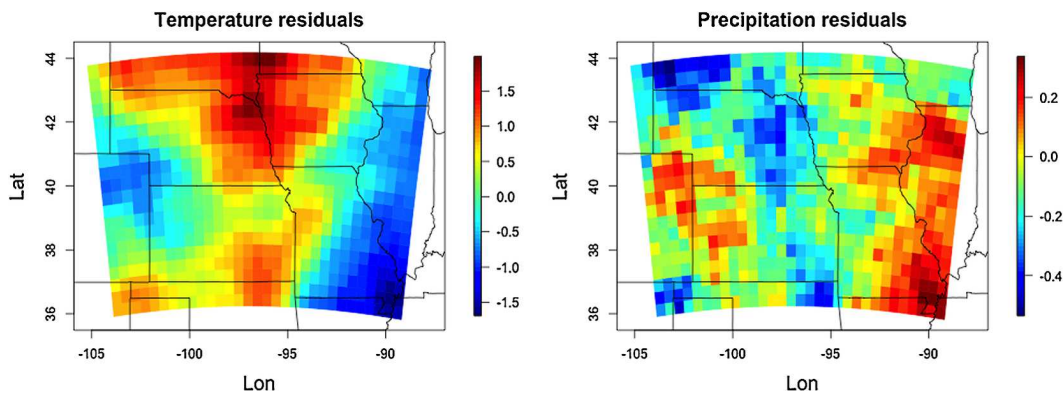}

\caption{Example residuals from 1989 after removing a spatially
varying mean
from NCEP-driven ECP2 regional climate model run
for the variables of average summer temperature and precipitation.
Units are degrees Celsius for temperature and centimeters for precipitation.}\label{fig:temp.prec}
\end{figure*}
%
%%%%%%%%%%%%%%%%%%%%%%%%%%%%%%%%%%%%%%%%%%%%%%%%%%%%%%%%%%%%%%%%

%%%%%%%%%%%%%%%%%%%%%%%%%%%%%%%%%%%%%%%%%%%%%%%%%%%%%%%%%%%%%%%%%%%%%%%%
%s5.3 #&#
\subsection{Cross-Covariance Functions on the Sphere}

Many multivariate datasets from environmental and climate sciences are
collected over large portions
of the Earth, for example, by satellites and, therefore,
cross-covariance functions on the sphere $\Sp^2$ in $\R^3$ are in need.
Consider a multivariate process on the sphere for which the $i$th
variable is described by
$Z_i(L,l)$, $i=1,\ldots,p$, with $L$ denoting latitude and $l$
denoting longitude.
\citet{Jun11} constructed cross-covariance functions by applying
differential operators with respect to
latitude and longitude to the process on the sphere. Furthermore, \citet{Jun11} studied nonstationary models
of cross-covariances with respect to latitude, so-called axially
symmetric, and longitudinally irreversible
cross-covariance functions for which
\begin{eqnarray}
&&\cov \bigl\{Z_i(L_1,l_1),Z_j(L_2,l_2)
\bigr\}\nonumber\\
&&\quad \neq\cov \bigl\{ Z_i(L_1,l_2),Z_j(L_2,l_1)
\bigr\},\nonumber
\end{eqnarray}
$(L_1,l_1)\in\Sp^2,
(L_2,l_2)\in\Sp^2$. All the models described in \citet{Jun11} are valid for the chordal
distance, that is, the Euclidean distance in $\R^3$
between points on~$\Sp^2$.
\citet{CasGen14} relaxed the assumption of axial symmetry
for univariate random fields on the sphere
and the extension of their work to multivariate random fields on the
sphere remains an open problem.
\citet{Gne13} provided a very thorough study of positive definite
functions on
a sphere that can be used as covariances. \citet{DuMaLi13} developed a
characterization of isotropic and continuous variogram matrix functions on
the sphere, extending some of the ideas of \citet{Ma12} who characterized
continuous and isotropic
covariance matrix functions on the sphere using Gegenbauer polynomials.
Because the great circles are the geodesics on the sphere, they are the
natural metric to measure distances in this context. \citet{PorBevGen} developed cross-covariance functions
of the great circle distances on the sphere. In particular, they
studied multivariate Mat\'ern models as
functions of the great circle distance on the sphere. Recently, \citet{Jun14} developed nonstationary
Mat\'ern cross-covariance models whose smoothness parameters vary over
space and with large-scale
nonstationarity obtained with the aforementioned differential operators.

%%%%%%%%%%%%%%%%%%%%%%%%%%%%%%%%%%%%%%%%%%%%%%%%%%%%%%%%%%%%%%%%%%%%%%%%
%s6 #&#
\section{Data Examples}
%%%%%%%%%%%%%%%%%%%%%%%%%%%%%%%%%%%%%%%%%%%%%%%%%%%%%%%%%%%%%%%%%%%%%%%%

We illustrate a selection of the above cross-covari\-ance models on two
data examples.
First, a set of reanalysis climate model output that represents
spatially gridded data.
Second, a set of observational temperature data that illustrates
spatially irregularly located data.

%%%%%%%%%%%%%%%%%%%%%%%%%%%%%%%%%%%%%%%%%%%%%%%%%%%%%%%%%%%%%%%%
%

%%%%%%%%%%%%%%%%%%%%%%%%%%%%%%%%%%%%%%%%%%%%%%%%%%%%%%%%%%%%%%%%%%%%%%%%
%s6.1 #&#
\subsection{Climate Model Output Data}

The specific reanalysis dataset in use is a
National Centers for Environmental Protection-driven (NCEP)
run of the updated Experimental Climate Prediction Center
(ECP2) model, which was originally run as part of the North American Regional
Climate Change Assessment Program (NARCCAP)
climate modeling experiment \citep{Meaetal09}. Reanalysis data can
be thought of as an estimate of the true state of the atmosphere for a
given period.
The variables we use are average summer temperature and cube-root precipitation
(summer being comprised of June, July and August; JJA) over a region of
the midwest United States
that is largely an agricultural region with relatively constant terrain.
The cube-root transformation reduces skewness in the precipitation
output and
brings the distribution closer to Gaussian.
For each grid cell, we calculate a pointwise spatially varying mean
as the arithmetic average of all 24 years of model output from 1981 through
2004. The data considered then are 24 years of residuals, having removed
this spatially varying mean from each year's reanalysis output for the two
variables of temperature and cube-root precipitation. The residuals are
assumed to be
independent between years, and are additionally assumed to be realizations
from a mean zero bivariate Gaussian process (both assumptions are
supported by
exploratory analysis).
%
%%%%%%%%%%%%%%%%%%%%%%%%%%%%%%%%%%%%%%%%%%%%%%%%%%%%%%%%%%%%%%%%%%%%%%%%%%%%%%%%%%%%%
%
%
%
%t1 #&#
\begin{table*}[b]
\tablewidth=\textwidth
\tabcolsep=0pt
\caption{Maximum likelihood estimates of parameters for full and
parsimonious bivariate Mat\'ern models, applied to the
NARCCAP model data. Units are degrees Celsius for temperature,
centimeters for precipitation, and kilometers for distances}\label{tab:estimates}
\begin{tabular*}{\textwidth}{@{\extracolsep{\fill}}lcccccccc@{}}
\hline
\textbf{Model} & $\boldsymbol{\sigma}_{\boldsymbol{T}}$ & $\boldsymbol{\sigma}_{\boldsymbol{P}}$
& $\boldsymbol{\nu}_{\boldsymbol{T}}$ & $\boldsymbol{\nu}_{\boldsymbol{P}}$ & $\boldsymbol{1/a_T}$ &
$\boldsymbol{1/a_P}$ & $\boldsymbol{1/a_{\mathit{TP}}}$ & $\boldsymbol{\rho}_{\boldsymbol{\mathit{TP}}}$  \\
\hline
Full & 1.63 & 0.19 & 1.31 & 0.55 & 384.3 & 361.6 & 420.1 & $-0.60$
 \\
Parsimonious & 1.61 & 0.19 & 1.33 & 0.54 & 367.1 & -- & -- & $-0.49$
\\
\hline
\end{tabular*}
\end{table*}

Figure~\ref{fig:temp.prec} contains an example set of reanalysis residuals
for the year 1989. By eye, it appears that temperature residuals are smoother
over space, while precipitation is apparently rougher, while both seem
to have
similar correlation length scales. The two variables are strongly negatively
correlated, with an empirical correlation coefficient of $-0.67$. This
situation,
with negative and strong cross-correlation and both variables exhibiting
distinct levels of smoothness, provides numerous challenges to available
cross-correlation models. Call $T(\bs,t)$ and $P(\bs,t)$ the
temperature and
precipitation residual at location $\bs$ in year $t$, respectively (recalling
that, although indexed by year, the processes are viewed as
temporally-independent).

Of the above models, we compare six to an independence assumption, that is,
where temperature and precipitation residuals are assumed to be independent;
for the independence model, each variable is assumed to follow a Mat\'ern
covariance, and parameters are estimated by maximum likelihood.
The first nontrivial bivariate model is the parsimonious Mat\'ern,
whose parameters
we estimate by maximum likelihood. The second model is a nearly full bivariate
Mat\'ern, where we set the cross-covariance smoothness $\nu_{\mathit{TP}}$, $T$
representing temperature and $P$ precipitation, to be the arithmetic average
of the marginal smoothnesses. For the full bivariate Mat\'ern, we set
marginal parameters to be those of the independence model, and conditional
on these, estimate the remaining cross-covariance length scale $a_{\mathit{TP}}$ and
cross-correlation coefficient $\rho_{\mathit{TP}}$ by maximum likelihood. We
additionally consider two variations on the bivariate parsimonious Mat\'ern,
one using a lagged covariance of \citet{LiZha11} (see Section~5.1),
and a nonstationary Mat\'ern with spatially varying variances for both
variables. Spatially varying variances are estimated empirically
at each grid cell, and conditional on these, the remaining parameters
are estimated
by maximum likelihood. We also consider a linear model of coregionalization,
\begin{eqnarray*}
T(\bs,t) &=& a_{11} Z_1(\bs,t),
\\
P(\bs,t) &=& a_{12} Z_1(\bs,t) + a_{22}
Z_2(\bs,t),
\end{eqnarray*}
where $Z_1$ and $Z_2$ are independent mean zero
spatial processes with Mat\'ern covariances.
We opt for this formulation since temperature is expected to be smoother
than precipitation, and our goal is to preserve this feature within the
statistical model. Parameters are estimated by maximum likelihood. Finally,
we additionally consider two latent dimensional models. The first is
parameterized
by (\ref{latgenexample}), except without a nugget effect, and the
second is
built via\vspace*{-1pt}
\begin{eqnarray*}
T(\bs,t) &= &b_{11} Z(\bs,t) + b_{12} Z_1(
\bs,t),
\\
P(\bs,t) &=& b_{21} Z(\bs,t) + b_{22} Z_2(\bs,t),
\end{eqnarray*}
where $Z(\bs,t)$ has a latent dimensional covariance of the form
\begin{eqnarray*}
C(\bh)&=&\frac{1}{(\|\bxi_i-\bxi_j\|+1)^\beta} \exp \biggl\{ \frac
{-\alpha\|\bh\|^2}{(\|\bxi_i-\bxi_j\|+1)^{\beta
}} \biggr\},
\end{eqnarray*}
$\bh\in\R^2$, and $Z_1, Z_2$ are independent with Mat\'ern correlations. This choice for
$Z$ allows the temperature process to retain smoother behavior at the origin
than precipitation, whereas the model of (\ref{latgenexample}) forces
exponential-like
behavior at the origin.
%
%
%t2 #&#
\begin{table*}
\tablewidth=\textwidth
\tabcolsep=0pt
\caption{Comparison of log likelihood values and pseudo cross-validation
scores averaged over ten cross-validation replications
for various multivariate models on the NARCCAP model data residuals for
temperature (T) and precipitation (P)}\label{tab:comparison}
\begin{tabular*}{\textwidth}{@{\extracolsep{\fill}}lccccc@{}}
\hline
& \textbf{Log likelihood}
& \textbf{RMSE (}$\boldsymbol{T}$\textbf{)} & \textbf{CRPS (}$\boldsymbol{T}$\textbf{)}
& \textbf{RMSE (}$\boldsymbol{P}$\textbf{)} & \textbf{CRPS (}$\boldsymbol{P}$\textbf{)}  \\
\hline
Nonstationary parsimonious Mat\'ern & $53564.5$
& 0.168 & 0.084 & 0.085 & 0.047 \\
Parsimonious lagged Mat\'ern & $52563.7$
& 0.179 & 0.090 & 0.087 & 0.048 \\
Full Mat\'ern & $52560.1$
& 0.178 & 0.090 & 0.087 & 0.048 \\
Parsimonious Mat\'ern & $52556.9$
& 0.179 & 0.090 & 0.087 & 0.048 \\
Latent dimension & $52028.8$
& 0.180 & 0.091 & 0.088 & 0.049 \\
LMC & $51937.0$
& 0.179 & 0.091 & 0.090 & 0.050 \\
Independent Mat\'ern & $50354.5$
& 0.180 & 0.091 & 0.088 & 0.049 \\
Latent dimension of (\ref{latgenexample}) & $48086.3$
& 0.195 & 0.100 & 0.088 & 0.048 \\
\hline
\end{tabular*}
\end{table*}
%
%%%%%%%%%%%%%%%%%%%%%%%%%%%%%%%%%%%%%%%%%%%%%%%%%%%%%%%%%%%%%%%%%%%%%%%%%%%%%%%%%%%%%%

Table~\ref{tab:estimates} contains the parsimonious and full bivariate
Mat\'ern
parameter estimates. Note the smoothness parameter of the temperature
field is
approximately $1.3$, indicating a relatively smooth field, which
supports the
theoretical analysis of \citet{NorWanGen11}; on the other hand, precipitation
has a smoothness of approximately $0.55$, suggesting an exponential
model may
work well. Both variables have similar length scale parameters, which suggests
the assumptions of the parsimonious Mat\'ern model may be reasonable
for this
particular dataset. The cross-correlation coefficient is estimated to
be strongly
negative in both cases, with the full Mat\'ern slightly closer to the empirical
cross-correlation.

Table~\ref{tab:comparison} contains log likelihood values for the
various models
considered. Evidently, the parsimonious, full and parsimonious lagged
Mat\'ern
all have likelihood values on the same order, which are all superior to
the LMC,
independent Mat\'ern and latent dimensional models. We remark that,
given the
smooth nature of the temperature field, the latent dimensional model of
(\ref{latgenexample}) is not
expected to perform as well, as it fixes the smoothness of the temperature
field at $\nu=0.5$, while on the other hand the latent dimensional
model using
a shared process with squared exponential covariance performs nearly as well
as the Mat\'ern alternatives. The nonstationary extension of the parsimonious
Mat\'ern exhibits the largest log likelihood, improving the next best
model by over
1000. This suggests that the bivariate field indeed exhibits
nonstationarity, and
there may be other modeling improvements that can be explored with new
nonstationary cross-covariance developments.
%

%%%%%%%%%%%%%%%%%%%%%%%%%%%%%%%%%%%%%%%%%%%%%%%%%%%%%%%%%%%%%%%%%%%%%%%%%%%%%%%%%%%%%%

Finally, we perform a small pseudo cross-validation study. We hold out
the bivariate
model output at a randomly chosen $90\%$ of spatial locations consistent
over all time points.
We then co-krige the remaining $10\%$ (62 locations) to the held out
grid cells using parameter estimates based on the entire dataset.
As the residual process is assumed to be independent between years, co-kriging
is performed separately for each year.
Root mean squared error (RMSE) and the continuous ranked probability
score (CRPS) are
used to validate interpolation quality, averaged over all held out
locations and
years. We repeat this experiment ten times for different randomly
chosen sets of
held out spatial locations and average the resulting scores; the
results are displayed in Table~\ref{tab:comparison}. Generally
speaking, all models are effectively equivalent
in terms of predictive ability, except for the nonstationary extension
to the
parsimonious Mat\'ern, which appears to improve both predictive
quantities for
temperature especially. Perhaps surprisingly, the independent Mat\'ern
performs as well for interpolation, although this has not been the case with
all datasets \citep{GneKleSch10}.

%%%%%%%%%%%%%%%%%%%%%%%%%%%%%%%%%%%%%%%%%%%%%%%%%%%%%%%%%%%%%%%%%%%%%%%%
%s6.2 #&#
\subsection{Observational Temperature Data}

The second example we consider is a bivariate minimum and maximum temperature
observational dataset. Observations are available at stations that are part
of the United States Historical Climatology Network \citep{PetVos97}
over the state of
Colorado. Stations in the USHCN form the highest quality observational
climate network in the United States; observations are subject to rigorous
quality control.

We consider bivariate daily temperature residuals (i.e., having removed
the state-wide mean) on September 19, 2004, a day which has good
network coverage
with observations being available at 94 stations. Exploratory Q--Q plots suggest
the residuals are well modeled marginally as Gaussian processes; we suppose
the bivariate process is a realization from a bivariate Gaussian
process with
zero mean.

%%%%%%%%%%%%%%%%%%%%%%%%%%%%%%%%%%%%%%%%%%%%%%%%%%%%%%%%%%%%%%%%%%%%%%%%%%%%%%%%%%%%%%
%
%
%t3 #&#
\begin{table*}
\tablewidth=\textwidth
\tabcolsep=0pt
\caption{Comparison of log likelihood values and pseudo cross-validation
scores averaged over 100 cross-validation replications
for various multivariate models on the USHCN observed temperature
residuals for maximum temperature (max) and minimum temperature (min)}\label{tab:comparison2}
\begin{tabular*}{\textwidth}{@{\extracolsep{\fill}}lccccc@{}}
\hline
& \textbf{Log likelihood}
& \textbf{RMSE (min)} & \textbf{CRPS (min)} & \textbf{RMSE (max)} & \textbf{CRPS (max)}
\\
\hline
Parsimonious lagged Mat\'ern & $-414.0$%$-327.6$
& 3.18 & 1.83 & 3.14 & 1.79 \\
Parsimonious Mat\'ern & $-414.9$%$-328.5$
& 3.22 & 1.85 & 3.16 & 1.80 \\
LMC & $-415.7$%$-329.3$
& 3.22 & 1.85 & 3.16 & 1.80 \\
Latent dimension & $-416.2$%$-329.8$
& 3.23 & 1.86 & 3.18 & 1.81 \\
Latent dimension of (\ref{latgenexample}) & $-419.1$%$-332.7$
& 3.24 & 1.86 & 3.17 & 1.81 \\
Independent Mat\'ern & $-427.6$%$-341.2$
& 3.41 & 1.94 & 3.35 & 1.91 \\
\hline
\end{tabular*}
\end{table*}

We entertain the same set of bivariate models as in the previous
example subsection.
Due to the fact that the data are observational, we augment each process'
covariance with a nugget effect. We begin by estimating
the independent Mat\'ern model separately for both minimum and maximum
temperature
residuals by maximum likelihood. Since the nugget effect is tied to
marginal process behavior, we fix the estimated nugget effects at their
marginal estimates, and estimate all other covariance parameters from the
remaining bivariate models by maximum likelihood, conditional on these marginal
nugget estimates. We remove both the bivariate Mat\'ern and nonstationary
model from consideration, as these are both difficult to estimate given a
single realization of the spatial process.
%
%%%%%%%%%%%%%%%%%%%%%%%%%%%%%%%%%%%%%%%%%%%%%%%%%%%%%%%%%%%%%%%%%%%%%%%%%%%%%%%%%%%%%%
%

%
%%%%%%%%%%%%%%%%%%%%%%%%%%%%%%%%%%%%%%%%%%%%%%%%%%%%%%%%%%%%%%%%%%%%%%%%%%%%%%%%%%%%%%

On top of comparing in sample log likelihood values, we additionally consider
a pseudo cross-validation study, leaving out a randomly selected $25\%$ of
locations, and co-krige the remaining bivariate observations to these
held out
locations. This pseudo cross-validation procedure is repeated 100
times, and
Table~\ref{tab:comparison2} contains the averaged scores from this study.
Contrasting with the results of the NARCCAP example, we now see the predictive
benefit of considering multivariate second-order structures. Generally,
predictive RMSE and CRPS are improved by between 6--7\% when
co-kriging using the
parsimonious lagged Mat\'ern, as compared to marginally kriging each variable.
A potential explanation for the improvement here as compared to the NARCCAP
example is that in the current study, the observations are subject to
measurement
error, and thus the greater uncertainty in estimating the bivariate surface
is more readily quantified using an appropriate bivariate covariance model.

%%%%%%%%%%%%%%%%%%%%%%%%%%%%%%%%%%%%%%%%%%%%%%%%%%%%%%%%%%%%%%%%%%%%%%%%
%s7 #&#
\section{Discussion}\label{sec:end}
%%%%%%%%%%%%%%%%%%%%%%%%%%%%%%%%%%%%%%%%%%%%%%%%%%%%%%%%%%%%%%%%%%%%%%%%

%%%%%%%%%%%%%%%%%%%%%%%%%%%%%%%%%%%%%%%%%%%%%%%%%%%%%%%%%%%%%%%%%%%%%%%%
%s7.1 #&#
\subsection{Specialized Cross-Covariance Functions}

The models introduced so far cover the broad majority of usual datasets
requiring
multivariate models. However, specialized scenarios sometimes arise,
and call
for novel developments. For instance, some constructions involve
modeling variables
that exhibit long range dependence. \citet{Ma11N3} examined a construction
for all
variables having long or short range dependence utilizing univariate variograms;
and \citet{Ma11N1} explored the relationship between multivariate
covariances and
variograms. \citet{KlePor} derived a nonstationary
construction that allows
individual variables to be a spatially varying mixture of short and
long range
dependence, as well as having substantial cross-correlation between variables
(with possibly opposing short/long range dependence); their
construction is a
special case of a multivariate generalization of the univariate Cauchy class
of covariance \citep{GneSch04}.
\citet{HriPor} defined the multivariate analogue of
Spartan Gibbs random fields, obtained through using Hamiltonian functionals.

\citet{Ma11N2} also studied various approaches to produce valid
cross-covariance functions
based on differentiation of univariate covariance functions and on
scale mixtures
of covariance matrix functions. Alternatively, \citet{Ma11N4} provided
constructions of
variogram matrix functions, and \citet{DuMa12} introduced an approach to
building variogram matrix functions based on a univariate variogram model.

We close this section by pointing out a recent novel approach to
generating valid
matrix covariances by considering stochastic partial differential equations
(SPDEs); \citet{Huetal} used systems of SPDEs to simultaneously model
temperature and humidity, yielding computationally efficient means to analysis
by approximating a Gaussian random field by a Gaussian Markov random field.

%%%%%%%%%%%%%%%%%%%%%%%%%%%%%%%%%%%%%%%%%%%%%%%%%%%%%%%%%%%%%%%%%%%%%%%%
%s7.2 #&#
\subsection{Spatio-Temporal Cross-Covariance Functions}\label{sec:spatiotemp}

So far, the cross-covariance models that we described were aimed at
spatial multivariate random fields. When adding the time dimension,
the resulting spatio-temporal multivariate random field, $\bZ(\bs,t)$,
has stationary cross-covariance functions $C_{ij}(\bh,u)$, where $u$ denotes
a time lag. All the previous spatial cross-covariance models can be
straightforwardly extended to the spatio-temporal setting, for example,
\citet{RouWac90},
\citet{Choetal09}, \citet{BerGelHol10} and \citet{DeIetal13}, \citet{DeIPalPos13}
developed space--time versions of the linear model of coregionalization.
\citet{GelBanGam05} used a dynamic approach for multivariate space--time
data using coregionalization.

Based on the concept of latent dimensions described in Section~\ref{sec:latdim}, \citet{ApaGen10} have extended a class of
spatio-temporal
covariance functions for univariate random fields due to \citet{Gne02} to the multivariate setting. Specifically, if $\varphi_1(t)$,
$t\geq0$, is
a completely monotone function and $\psi_1(t),\allowbreak  \psi_2(t)$, $t\geq0$,
are positive functions with completely monotone derivatives, then
%
%
%e13 #&#
\begin{eqnarray}\label
{eq:apagenhuv}
C(\bh,u,\bv)&=&\frac{\sigma^2}{ [\psi_1\{u^2/\psi_2(\|\bv\|
^2)\} ]^{d/2} \{\psi_2(\|\bv\|^2)\}^{1/2}}\nonumber\\
[-8pt]\\[-8pt]
&&{}\cdot\varphi_1 \biggl[ \frac
{\|\bh\|^2}{\psi_1\{u^2/\psi_2(\|\bv\|^2)\}}
\biggr], \nonumber
\end{eqnarray}
is a valid stationary covariance function on $\R^{d+1+k}$ that can be used
to model cross-covariance functions with $\bv=\bxi_i-\bxi_j$.
When $\psi_2(t)\equiv1$, Gneiting's class is retrieved. The case $\bv
=\mathbf{0}$
yields a common spatio-temporal covariance function for
each variable that can be made different through a LMC-type
construction. Also
judicious choices of the functions in (\ref{eq:apagenhuv}) allow one
to control
nonseparability between space and time, between space and variables,
and between
time and variables; see \citet{ApaGen10}
for various illustrative examples.

To further introduce asymmetry in spatio-temporal cross-covariance
functions, \citet{ApaGen10} have proposed two approaches
based on latent dimensions. Using the notation of Section~\ref{sec:latdim}, the first type of asymmetric spatio-temporal
cross-covariance is
%
%
%e14 #&#
\begin{eqnarray}\label{asym1}
\quad C_{ij}^{a}(\bh,u)=C \bigl(\bh,u-\blambda_\xi^{\mathrm{T}}(
\bxi_i-\bxi_j),\bxi_i-\bxi_j
\bigr),
\end{eqnarray}
$\bh\in\R^d, u \in\R$, where $C$ is a valid covariance function on $\R^{d+k}$ of a univariate
random field and $\blambda_\xi\in\R^k$, $1\leq k \leq p$, controls
the delay in time
that creates asymmetry. There is no time delay if and only if $\blambda
_\xi=\mathbf{0}$ or $i=j$. The second type of asymmetric
spatio-temporal cross-covariance is
%
%
%e15 #&#
\begin{eqnarray}\label{asym2}
\quad C_{ij}^{a}(\bh,u)=C(\bh-\bgamma_h u,u,
\bxi_i-\bxi_j-\bgamma_\xi u),
\end{eqnarray}
$\bh\in\R^d, u \in\R$, where the velocity vectors $\bgamma_h \in\R^d$ and $\bgamma_\xi\in
\R^k$ are responsible for the lack of symmetry. When $u\neq0$, this
model is spatially anisotropic. Combinations of models (\ref{asym1})
and (\ref{asym2}) are possible.

%%%%%%%%%%%%%%%%%%%%%%%%%%%%%%%%%%%%%%%%%%%%%%%%%%%%%%%%%%%%%%%%%%%%%%%%
%s7.3 #&#
\subsection{Physics-Constrained Cross-Covariance Functions}

Especially for geophysical processes, often there are physical
constraints on a
system of variables that must be obeyed by any stochastic model. For instance,
\citet{Bue72} explored valid covariance models for geostrophic wind that must
satisfy physical relationships for isotropic geophysical flow including
geopotential,
longitudinal wind components and transverse wind components.

In a similar vein, a number of physical processes, especially
in fluid dynamics, involve fields with specialized restrictions such as
being divergence free. \citet{SchSch12} developed matrix-valued
covariance functions for divergence-free and curl-free random vector fields,
which are based on combinations of derivatives of a specified variogram
and extend earlier work by \citet{NarWar94}.

\citet{ConAni13} introduced a framework for building valid
matrix-valued
covariance functions when the constituent processes have known physical
constraints
relating their behavior. By approximating a nonlinear physical relationship
between variables through series expansions and closures, the authors develop
physically-based matrix covariance classes.
They explored large-scale geostrophic wind as a case study,
and illustrated that physically motivated cross-correlation models can
substantially
outperform independence models.

\citet{NorWanGen11} studied spatio-temporal correlations for temperature
fields arising from simple energy-balance climate models, that is,
white-noise-driven damped diffusion equations. The resulting spatial
correlation on
the plane is of Mat\'ern type with smoothness parameter $\nu=1$,
although rougher
temperature fields are expected due to terrain irregularities for example.
Derivations for temperature fields on a uniform sphere were presented
as well.
Whether these results can be extended to other variables
such as pressure and wind fields, and possibly lead to Mat\'ern cross-covariance
models of type (\ref{eq:mm}), is an open question.

%%%%%%%%%%%%%%%%%%%%%%%%%%%%%%%%%%%%%%%%%%%%%%%%%%%%%%%%%%%%%%%%%%%%%%%%
%s7.4 #&#
\subsection{Open Problems}

Finally, there are many open problems that call for more research. The most
fundamental question is the theoretical characterization of the
allowable classes
of multivariate covariances. For instance, given two marginal covariances,
what is the valid class of possible cross-covariances that still
results in a
nonnegative definite structure? Such a characterization is an unsolved problem.
Additional to characterization, the companion theoretical question is
the utility
of cross-covariance models. Given the two data examples in this review,
a natural
question is: for the purposes of co-kriging, in what situations are the
use of
nontrivial cross-covariances beneficial? Although it is traditional to focus
on kriging and co-kriging in the geostatistical literature, we wish to
additionally emphasize the utility of these models for simulation of
multivariate
random fields. Indeed, without flexible cross-covariance models, it is
impossible to simulate multiple fields with nontrivial dependencies.

The power exponential class of covariances is a useful marginal class of
covariances, but
to the best of our knowledge, a characterization of parameters for the validity
of the multivariate version
\begin{eqnarray}
\rho_{ij}(\bh)=\beta_{ij}\exp \biggl\{ - \biggl(
\frac{\|\bh\|}{\phi
_{ij}} \biggr)^{\kappa_{ij}} \biggr\}, \quad \bh\in\R^d,
\nonumber
\end{eqnarray}
is not known. Although we believe that the multivariate Mat\'ern model
(\ref{eq:mm}) has more flexibility, this is
still an interesting question, especially as this set of covariances
requires no
calculations involving Bessel functions.

The extension of spatial extremes to the case of multiple variables has
not been explored yet except for the recent
proposal of \citet{GenPadSan} who considered multivariate max-stable
spatial processes. The aim
of that research is to describe the behavior of extreme events of
several variables across space,
such as extreme rainfall and extreme temperature for example.
This requires flexible and physically-realistic cross-covariance models
and therefore the families described herein may play an important role
for such applications.

Recently, there has been some new interest in other types of random
fields than
the usual Gaussian case. Mittag--Leffler fields contain the Gaussian
case as a
subset, but are specified in terms of an infinite series expansion that is
unwieldy for applications \citep{Ma}. Another option is a multivariate
extension of the Student's $t$ distribution, a $t$-vector distribution \citep{Ma13};
these seem to be more promising for applications, and some exploration of
the utility of these types of models is called for. Finally, hyperbolic
vector random fields contain the Student's $t$ as a limiting case, although
model interpretation, estimation and implementation remain unexplored
(\cite{Duetal12}).

There is also a need for valid multivariate cross-covariance functions
for spatial
data on a lattice. Although one can apply any of the models mentioned in
this manuscript to lattice data, the extension of univariate Markov
random field
models is another route.
For instance, \citet{GelVou03} have studied proper
multivariate conditional autoregressive models.
\citet{DanZhoZou06} proposed a class of conditionally specified
space--time models for multivariate
processes geared to situations where there is a sparse spatial coverage
of one of the processes and
a much more dense coverage of the other processes. This is motivated by
an application to particulate matter and ozone data.
\citet{SaiCre07} also developed Markov random field models for
multivariate lattice data.

Many additional open questions remain, including theoretical
development of
estimation in the multivariate context \citep{PasZha06}.
\citet{VarWarMye99} looked at the relationship between support
size and relationship between variables, but relatively few have
explored this
phenomenon in the multivariate case.
Finally, there is a need to better understand and explore the
intimate connection between multivariate spline smoothers, co-kriging and
multivariate numerical analysis (\cite*{BeazuCSch11}; \cite*{Fus08}; \cite*{NarWar94};
\citeauthor{ReiBur07}, \citeyear{ReiBur07}).

%\begin{appendix}
%\section{}
%\end{appendix}

%\section*{Acknowledgments}

%\begin{supplement}[id=suppA]
%\slink[doi]{10.1214/00-STSXXXXSUPP} %[doi,text={...}] - jei reikia
%suskaldyti doi
%\sfilename{stsXXXX\_supp.pdf}
%\end{supplement}

% imsref loaded by jurgita.kaciuliene, 2014-07-10 09:28:20
% imsref loaded by jurgita.kaciuliene, 2014-07-10 11:18:08
%
% imsref loaded by jurgita.kaciuliene, 2014-07-10 12:21:02
% imsref loaded by jurgita.kaciuliene, 2014-07-10 12:31:43
% imsref loaded by akundreckaite, 2015-05-11 10:16:10


\begin{thebibliography}{123}

%b1 ###
\bibitem[\protect\citeauthoryear{Almeida and Journel}{1994}]{AlmJou94}
%
\begin{barticle}[auto:STB|2014/06/18|12:29:53]
\bauthor{\bsnm{Almeida},~\bfnm{A.~S.}\binits{A.~S.}} \AND
\bauthor{\bsnm{Journel},~\bfnm{A.~G.}\binits{A.~G.}}
(\byear{1994}).
\btitle{Joint simulation of multiple variables with a Markov-type
coregionalization model}.
\bjournal{Math. Geol.}
\bvolume{26}
\bpages{565--588}.
\end{barticle}
%
\bptok{imsref}%
\endbibitem

%b2 ###
\bibitem[\protect\citeauthoryear{Alonso-Malaver, Porcu and
Giraldo}{2013}]{AloPorGir}
%
\begin{bmisc}[auto:STB|2014/06/18|12:29:53]
\bauthor{\bsnm{Alonso-Malaver},~\bfnm{C.}\binits{C.}},
\bauthor{\bsnm{Porcu},~\bfnm{E.}\binits{E.}} \AND
\bauthor{\bsnm{Giraldo},~\bfnm{R.}\binits{R.}}
(\byear{2013}).
\bhowpublished{Multivariate versions of walks through dimensions and
Schoenberg measures. Technical report, Univ. Tecnica Federico
Santa Maria, Valparaiso, Chile}.
\end{bmisc}
%
\bptok{imsref}%
% NOT OUTPUTED:
% sortkey = Alonso
% howpublished = (2013a), ``Multivariate versions of walks through
%dimensions and Schoenberg measures, '' \textit{Technical Report
%University Federico Santa Maria}
\endbibitem

%b3 ###
\bibitem[\protect\citeauthoryear{Alonso-Malaver, Porcu and
Giraldo}{2015}]{autokey3}
%
\begin{barticle}[auto:STB|2014/06/18|12:29:53]
\bauthor{\bsnm{Alonso-Malaver},~\bfnm{C.}\binits{C.}},
\bauthor{\bsnm{Porcu},~\bfnm{E.}\binits{E.}} \AND
\bauthor{\bsnm{Giraldo},~\bfnm{R.}\binits{R.}}
(\byear{2015}).
\btitle{Multivariate and multiradial Schoenberg measures with
their dimension walk}.
\bjournal{J. Multivariate Anal.}
\bvolume{133}
\bpages{251--265}.
\bid{mr={3282029}}
\end{barticle}
%
\bptok{imsref}%
% NOT OUTPUTED:
% sortkey = Alonso
% howpublished = Alonso-Malaver, C. Porcu, E., and Giraldo, R. (2013b),
%``Multivariate and multiradial Schoenberg measures with their
%dimension walk, '' \textit{Technical Report University Federico Santa
%Maria}
\endbibitem

%b4 ###
\bibitem[\protect\citeauthoryear{{\'A}lvarez, Rosasco and
Lawrence}{2012}]{AlvRosLaw12}
%
\begin{barticle}[auto:STB|2014/06/18|12:29:53]
\bauthor{\bsnm{{\'A}lvarez},~\bfnm{M.~A.}\binits{M.~A.}},
\bauthor{\bsnm{Rosasco},~\bfnm{L.}\binits{L.}} \AND
\bauthor{\bsnm{Lawrence},~\bfnm{N.~D.}\binits{N.~D.}}
(\byear{2012}).
\btitle{Kernels for vector-valued functions: A review}.
\bjournal{Found. Trends Mach. Learn.}
\bvolume{3}
\bpages{195--266}.
\end{barticle}
%
\bptok{imsref}%
\endbibitem

%b5 ###
\bibitem[\protect\citeauthoryear{Apanasovich and Genton}{2010}]{ApaGen10}
%
\begin{barticle}[mr]
\bauthor{\bsnm{Apanasovich},~\bfnm{Tatiyana~V.}\binits{T.~V.}} \AND
\bauthor{\bsnm{Genton},~\bfnm{Marc~G.}\binits{M.~G.}}
(\byear{2010}).
\btitle{Cross-covariance functions for multivariate random fields
based on latent dimensions}.
\bjournal{Biometrika}
\bvolume{97}
\bpages{15--30}.
\bid{doi={10.1093/biomet/asp078}, issn={0006-3444}, mr={2594414}}
\end{barticle}
%
\bptok{imsref}%
% NOT OUTPUTED:
% issn = 0006-3444
% url = http://dx.doi.org/10.1093/biomet/asp078
% number = 1
% coden = BIOKAX
% fjournal = Biometrika
\endbibitem

%b6 ###
\bibitem[\protect\citeauthoryear{Apanasovich, Genton and
Sun}{2012}]{ApaGenSun12}
%
\begin{barticle}[mr]
\bauthor{\bsnm{Apanasovich},~\bfnm{Tatiyana~V.}\binits{T.~V.}},
\bauthor{\bsnm{Genton},~\bfnm{Marc~G.}\binits{M.~G.}} \AND
\bauthor{\bsnm{Sun},~\bfnm{Ying}\binits{Y.}}
(\byear{2012}).
\btitle{A valid {M}at\'ern class of cross-covariance functions for
multivariate random fields with any number of components}.
\bjournal{J. Amer. Statist. Assoc.}
\bvolume{107}
\bpages{180--193}.
\bid{doi={10.1080/01621459.2011.643197}, issn={0162-1459}, mr={2949350}}
\end{barticle}
%
\bptok{imsref}%
% NOT OUTPUTED:
% issn = 0162-1459
% url = http://dx.doi.org/10.1080/01621459.2011.643197
% number = 497
% coden = JSTNAL
% fjournal = Journal of the American Statistical Association
\endbibitem

%b7 ###
\bibitem[\protect\citeauthoryear{Beatson, zu~Castell and Schr{\"
o}dl}{2011}]{BeazuCSch11}
%
\begin{barticle}[mr]
\bauthor{\bsnm{Beatson},~\bfnm{R.~K.}\binits{R.~K.}},
\bauthor{\bsnm{zu Castell},~\bfnm{W.}\binits{W.}} \AND
\bauthor{\bsnm{Schr{\"o}dl},~\bfnm{S.~J.}\binits{S.~J.}}
(\byear{2011}).
\btitle{Kernel-based methods for vector-valued data with correlated
components}.
\bjournal{SIAM J. Sci. Comput.}
\bvolume{33}
\bpages{1975--1995}.
\bid{doi={10.1137/090758076}, issn={1064-8275}, mr={2831042}}
\end{barticle}
%
\bptok{imsref}%
% NOT OUTPUTED:
% issn = 1064-8275
% url = http://dx.doi.org/10.1137/090758076
% number = 4
% coden = SJOCE3
% fjournal = SIAM Journal on Scientific Computing
\endbibitem

%b8 ###
\bibitem[\protect\citeauthoryear{Berrocal, Gelfand and
Holland}{2010}]{BerGelHol10}
%
\begin{barticle}[mr]
\bauthor{\bsnm{Berrocal},~\bfnm{Veronica~J.}\binits{V.~J.}},
\bauthor{\bsnm{Gelfand},~\bfnm{Alan~E.}\binits{A.~E.}} \AND
\bauthor{\bsnm{Holland},~\bfnm{David~M.}\binits{D.~M.}}
(\byear{2010}).
\btitle{A bivariate space--time downscaler under space and time misalignment}.
\bjournal{Ann. Appl. Stat.}
\bvolume{4}
\bpages{1942--1975}.
\bid{doi={10.1214/10-AOAS351}, issn={1932-6157}, mr={2829942}}
\end{barticle}
%
\bptok{imsref}%
% NOT OUTPUTED:
% issn = 1932-6157
% url = http://dx.doi.org/10.1214/10-AOAS351
% number = 4
% fjournal = The Annals of Applied Statistics
\endbibitem

%b9 ###
\bibitem[\protect\citeauthoryear{Bhat, Haran and Goes}{2010}]{autokey9}
%
\begin{bincollection}[mr]
\bauthor{\bsnm{Bhat},~\bfnm{K.}\binits{K.}},
\bauthor{\bsnm{Haran},~\bfnm{M.}\binits{M.}} \AND
\bauthor{\bsnm{Goes},~\bfnm{M.}\binits{M.}}
(\byear{2010}).
\btitle{Computer model calibration with multivariate spatial output: A case
study}.
In \bbooktitle{Frontiers of Statistical Decision Making and {B}ayesian Analysis}
(\beditor{\bfnm{Ming-Hui}\binits{M.-H.} \bsnm{Chen}},
\beditor{\bfnm{Dipak K.}\binits{D. K.} \bsnm{Dey}},
\beditor{\bfnm{Peter}\binits{P.} \bsnm{M{\"u}ller}},
\beditor{\bfnm{Dongchu}\binits{D.} \bsnm{Sun}} \AND
\beditor{\bfnm{Keying}\binits{K.} \bsnm{Ye}}, eds.)
\bpages{168--184}.
\bpublisher{Springer},
\blocation{New York}.
\bid{doi={10.1007/978-1-4419-6944-6}, mr={2766461}}
\end{bincollection}
%
\bptok{imsref}%
% NOT OUTPUTED:
% isbn = 978-1-4419-6943-9
% url = http://dx.doi.org/10.1007/978-1-4419-6944-6
% fpage = xxiv+631
\endbibitem

%b10 ###
\bibitem[\protect\citeauthoryear{Bochner}{1955}]{Boc55}
%
\begin{bbook}[mr]
\bauthor{\bsnm{Bochner},~\bfnm{Salomon}\binits{S.}}
(\byear{1955}).
\btitle{Harmonic Analysis and the Theory of Probability}.
\bpublisher{Univ. California Press},
\blocation{Berkeley and Los Angeles}.
\bid{mr={0072370}}
\end{bbook}
%
\bptok{imsref}%
% NOT OUTPUTED:
% fpage = viii+176
\endbibitem

%b11 ###
\bibitem[\protect\citeauthoryear{Bornn, Shaddick and
Zidek}{2012}]{BorShaZid12}
%
\begin{barticle}[mr]
\bauthor{\bsnm{Bornn},~\bfnm{Luke}\binits{L.}},
\bauthor{\bsnm{Shaddick},~\bfnm{Gavin}\binits{G.}} \AND
\bauthor{\bsnm{Zidek},~\bfnm{James~V.}\binits{J.~V.}}
(\byear{2012}).
\btitle{Modeling nonstationary processes through dimension expansion}.
\bjournal{J. Amer. Statist. Assoc.}
\bvolume{107}
\bpages{281--289}.
\bid{doi={10.1080/01621459.2011.646919}, issn={0162-1459}, mr={2949359}}
\end{barticle}
%
\bptok{imsref}%
% NOT OUTPUTED:
% issn = 0162-1459
% url = http://dx.doi.org/10.1080/01621459.2011.646919
% number = 497
% coden = JSTNAL
% fjournal = Journal of the American Statistical Association
\endbibitem

%b12 ###
\bibitem[\protect\citeauthoryear{Bourgault and Marcotte}{1991}]{BouMar91}
%
\begin{barticle}[auto:STB|2014/06/18|12:29:53]
\bauthor{\bsnm{Bourgault},~\bfnm{G.}\binits{G.}} \AND
\bauthor{\bsnm{Marcotte},~\bfnm{D.}\binits{D.}}
(\byear{1991}).
\btitle{Multivariable variogram and its application to the linear
model of coregionalization}.
\bjournal{Math. Geol.}
\bvolume{23}
\bpages{899--928}.
\end{barticle}
%
\bptok{imsref}%
\endbibitem

%b13 ###
\bibitem[\protect\citeauthoryear{Brown, Le and Zidek}{1994}]{BroLeZid94}
%
\begin{barticle}[mr]
\bauthor{\bsnm{Brown},~\bfnm{Philip~J.}\binits{P.~J.}},
\bauthor{\bsnm{Le},~\bfnm{Nhu~D.}\binits{N.~D.}} \AND
\bauthor{\bsnm{Zidek},~\bfnm{James~V.}\binits{J.~V.}}
(\byear{1994}).
\btitle{Multivariate spatial interpolation and exposure to air pollutants}.
\bjournal{Canad. J. Statist.}
\bvolume{22}
\bpages{489--509}.
\bid{doi={10.2307/3315406}, issn={0319-5724}, mr={1321471}}
\end{barticle}
%
\bptok{imsref}%
% NOT OUTPUTED:
% issn = 0319-5724
% url = http://dx.doi.org/10.2307/3315406
% number = 4
% fjournal = The Canadian Journal of Statistics. La Revue Canadienne de
%Statistique
\endbibitem

%b14 ###
\bibitem[\protect\citeauthoryear{Buell}{1972}]{Bue72}
%
\begin{barticle}[auto:STB|2014/06/18|12:29:53]
\bauthor{\bsnm{Buell},~\bfnm{C.~E.}\binits{C.~E.}}
(\byear{1972}).
\btitle{Correlation functions for wind and geopotential on isobaric surfaces}.
\bjournal{J. Appl. Meteorol.}
\bvolume{11}
\bpages{51--59}.
\end{barticle}
%
\bptok{imsref}%
\endbibitem

%b15 ###
\bibitem[\protect\citeauthoryear{Calder}{2007}]{Cal07}
%
\begin{barticle}[mr]
\bauthor{\bsnm{Calder},~\bfnm{Catherine~A.}\binits{C.~A.}}
(\byear{2007}).
\btitle{Dynamic factor process convolution models for multivariate
space--time data with application to air quality assessment}.
\bjournal{Environ. Ecol. Stat.}
\bvolume{14}
\bpages{229--247}.
\bid{doi={10.1007/s10651-007-0019-y}, issn={1352-8505}, mr={2405328}}
\end{barticle}
%
\bptok{imsref}%
% NOT OUTPUTED:
% issn = 1352-8505
% url = http://dx.doi.org/10.1007/s10651-007-0019-y
% number = 3
% coden = EESTFM
% fjournal = Environmental and Ecological Statistics
\endbibitem

%b16 ###
\bibitem[\protect\citeauthoryear{Calder}{2008}]{Cal08}
%
\begin{barticle}[mr]
\bauthor{\bsnm{Calder},~\bfnm{Catherine~A.}\binits{C.~A.}}
(\byear{2008}).
\btitle{A dynamic process convolution approach to modeling ambient
particulate matter concentrations}.
\bjournal{Environmetrics}
\bvolume{19}
\bpages{39--48}.
\bid{doi={10.1002/env.852}, issn={1180-4009}, mr={2416543}}
\end{barticle}
%
\bptok{imsref}%
% NOT OUTPUTED:
% issn = 1180-4009
% url = http://dx.doi.org/10.1002/env.852
% number = 1
% fjournal = Environmetrics
\endbibitem

%b17 ###
\bibitem[\protect\citeauthoryear{Castruccio and Genton}{2014}]{CasGen14}
%
\begin{barticle}[auto:STB|2014/06/18|12:29:53]
\bauthor{\bsnm{Castruccio},~\bfnm{S.}\binits{S.}} \AND
\bauthor{\bsnm{Genton},~\bfnm{M.~G.}\binits{M.~G.}}
(\byear{2014}).
\btitle{Beyond axial symmetry: An improved class of models for global data}.
\bjournal{Stat}
\bvolume{3}
\bpages{48--55}.
\end{barticle}
%
\bptok{imsref}%
\endbibitem

%b18 ###
\bibitem[\protect\citeauthoryear{Choi et~al.}{2009}]{Choetal09}
%
\begin{barticle}[mr]
\bauthor{\bsnm{Choi},~\bfnm{Jungsoon}\binits{J.}},
\bauthor{\bsnm{Reich},~\bfnm{Brian~J.}\binits{B.~J.}},
\bauthor{\bsnm{Fuentes},~\bfnm{Montserrat}\binits{M.}} \AND
\bauthor{\bsnm{Davis},~\bfnm{Jerry~M.}\binits{J.~M.}}
(\byear{2009}).
\btitle{Multivariate spatial--temporal modeling and prediction of
speciated fine particles}.
\bjournal{J. Stat. Theory Pract.}
\bvolume{3}
\bpages{407--418}.
\bid{doi={10.1080/15598608.2009.10411933}, issn={1559-8608}, mr={2751608}}
\end{barticle}
%
\bptok{imsref}%
% NOT OUTPUTED:
% issn = 1559-8608
% url = http://dx.doi.org/10.1080/15598608.2009.10411933
% number = 2
% fjournal = Journal of Statistical Theory and Practice
\endbibitem

%b19 ###
\bibitem[\protect\citeauthoryear{Constantinescu and
Anitescu}{2013}]{ConAni13}
%
\begin{barticle}[mr]
\bauthor{\bsnm{Constantinescu},~\bfnm{Emil~M.}\binits{E.~M.}} \AND
\bauthor{\bsnm{Anitescu},~\bfnm{Mihai}\binits{M.}}
(\byear{2013}).
\btitle{Physics-based covariance models for {G}aussian processes with
multiple outputs}.
\bjournal{Int. J. Uncertain. Quantif.}
\bvolume{3}
\bpages{47--71}.
\bid{doi={10.1615/Int.J.UncertaintyQuantification.2012003722},
issn={2152-5080}, mr={3044910}}
\end{barticle}
%
\bptok{imsref}%
% NOT OUTPUTED:
% issn = 2152-5080
% url =
%http://dx.doi.org/10.1615/Int.J.UncertaintyQuantification.2012003722
% number = 1
% fjournal = International Journal for Uncertainty Quantification
\endbibitem

%b20 ###
\bibitem[\protect\citeauthoryear{Cram{\'e}r}{1940}]{Cra40}
%
\begin{barticle}[mr]
\bauthor{\bsnm{Cram{\'e}r},~\bfnm{Harald}\binits{H.}}
(\byear{1940}).
\btitle{On the theory of stationary random processes}.
\bjournal{Ann. of Math. (2)}
\bvolume{41}
\bpages{215--230}.
\bid{issn={0003-486X}, mr={0000920}}
\end{barticle}
%
\bptok{imsref}%
% NOT OUTPUTED:
% issn = 0003-486X
% fjournal = Annals of Mathematics. Second Series
\endbibitem

%b21 ###
\bibitem[\protect\citeauthoryear{Cressie}{1993}]{Cre93}
%
\begin{bbook}[mr]
\bauthor{\bsnm{Cressie},~\bfnm{Noel~A.~C.}\binits{N.~A.~C.}}
(\byear{1993}).
\btitle{Statistics for Spatial Data}.
\bseries{Wiley Series in Probability and Mathematical Statistics:
Applied Probability and Statistics}.
\bpublisher{Wiley},
\blocation{New York}.
\bid{mr={1239641}}
\end{bbook}
%
\bptok{imsref}%
% NOT OUTPUTED:
% isbn = 0-471-00255-0
% fpage = xxii+900
\endbibitem

%b22 ###
\bibitem[\protect\citeauthoryear{Cressie and Wikle}{1998}]{CreWik98}
%
\begin{barticle}[mr]
\bauthor{\bsnm{Cressie},~\bfnm{Noel}\binits{N.}} \AND
\bauthor{\bsnm{Wikle},~\bfnm{Christopher~K.}\binits{C.~K.}}
(\byear{1998}).
\btitle{The variance-based cross-variogram: You can add apples and oranges}.
\bjournal{Math. Geol.}
\bvolume{30}
\bpages{789--799}.
\bid{doi={10.1023/A:1021770324434}, issn={0882-8121}, mr={1646242}}
\end{barticle}
%
\bptok{imsref}%
% NOT OUTPUTED:
% issn = 0882-8121
% url = http://dx.doi.org/10.1023/A:1021770324434
% number = 7
% coden = MATGED
% fjournal = Mathematical Geology
\endbibitem

%b23 ###
\bibitem[\protect\citeauthoryear{Daley, Porcu and
Bevilacqua}{2015}]{DalPorBev}
%
\begin{bmisc}[auto:STB|2014/06/18|12:29:53]
\bauthor{\bsnm{Daley},~\bfnm{D.~J.}\binits{D.~J.}},
\bauthor{\bsnm{Porcu},~\bfnm{E.}\binits{E.}} \AND
\bauthor{\bsnm{Bevilacqua},~\bfnm{M.}\binits{M.}}
(\byear{2015}).
\bhowpublished{Classes of compactly supported correlation functions
for multivariate random fields.
\textit{Stoch. Environ. Risk Assess.} To appear}.
\end{bmisc}
%
\bptok{imsref}%
% NOT OUTPUTED:
% sortkey = Daley(2014
% howpublished = (2014), ``Classes of compactly supported correlation
%functions for multivariate random fields, '' technical report,
%Department of Mathematics, Universidad Tecnica Federico Santa Maria,
%Valparaiso, Chile
\endbibitem

%b24 ###
\bibitem[\protect\citeauthoryear{Daniels, Zhou and Zou}{2006}]{DanZhoZou06}
%
\begin{barticle}[mr]
\bauthor{\bsnm{Daniels},~\bfnm{Michael~J.}\binits{M.~J.}},
\bauthor{\bsnm{Zhou},~\bfnm{Zhigang}\binits{Z.}} \AND
\bauthor{\bsnm{Zou},~\bfnm{Hui}\binits{H.}}
(\byear{2006}).
\btitle{Conditionally specified space--time models for multivariate processes}.
\bjournal{J. Comput. Graph. Statist.}
\bvolume{15}
\bpages{157--177}.
\bid{doi={10.1198/106186006X100434}, issn={1061-8600}, mr={2269367}}
\end{barticle}
%
\bptok{imsref}%
% NOT OUTPUTED:
% issn = 1061-8600
% url = http://dx.doi.org/10.1198/106186006X100434
% number = 1
% fjournal = Journal of Computational and Graphical Statistics
\endbibitem

%b25 ###
\bibitem[\protect\citeauthoryear{De~Iaco, Palma and Posa}{2013}]{DeIPalPos13}
%
\begin{barticle}[mr]
\bauthor{\bsnm{De Iaco},~\bfnm{S.}\binits{S.}},
\bauthor{\bsnm{Palma},~\bfnm{M.}\binits{M.}} \AND
\bauthor{\bsnm{Posa},~\bfnm{D.}\binits{D.}}
(\byear{2013}).
\btitle{Prediction of particle pollution through spatio-temporal
multivariate geostatistical analysis: Spatial special issue}.
\bjournal{Adv. Stat. Anal.}
\bvolume{97}
\bpages{133--150}.
\bid{doi={10.1007/s10182-012-0199-0}, issn={1863-8171}, mr={3045764}}
\end{barticle}
%
\bptok{imsref}%
% NOT OUTPUTED:
% issn = 1863-8171
% url = http://dx.doi.org/10.1007/s10182-012-0199-0
% number = 2
% fjournal = AStA. Advances in Statistical Analysis. A Journal of the
%German Statistical Society
\endbibitem

%b26 ###
\bibitem[\protect\citeauthoryear{De~Iaco et~al.}{2013}]{DeIetal13}
%
\begin{barticle}[mr]
\bauthor{\bsnm{De Iaco},~\bfnm{S.}\binits{S.}},
\bauthor{\bsnm{Myers},~\bfnm{D.~E.}\binits{D.~E.}},
\bauthor{\bsnm{Palma},~\bfnm{M.}\binits{M.}} \AND
\bauthor{\bsnm{Posa},~\bfnm{D.}\binits{D.}}
(\byear{2013}).
\btitle{Using simultaneous diagonalization to identify a space--time
linear coregionalization model}.
\bjournal{Math. Geosci.}
\bvolume{45}
\bpages{69--86}.
\bid{doi={10.1007/s11004-012-9408-3}, issn={1874-8961}, mr={3018750}}
\end{barticle}
%
\bptok{imsref}%
% NOT OUTPUTED:
% issn = 1874-8961
% url = http://dx.doi.org/10.1007/s11004-012-9408-3
% number = 1
% fjournal = Mathematical Geosciences
\endbibitem

%b27 ###
\bibitem[\protect\citeauthoryear{Du and Ma}{2012}]{DuMa12}
%
\begin{barticle}[mr]
\bauthor{\bsnm{Du},~\bfnm{Juan}\binits{J.}} \AND
\bauthor{\bsnm{Ma},~\bfnm{Chunsheng}\binits{C.}}
(\byear{2012}).
\btitle{Variogram matrix functions for vector random fields with
second-order increments}.
\bjournal{Math. Geosci.}
\bvolume{44}
\bpages{411--425}.
\bid{doi={10.1007/s11004-011-9377-y}, issn={1874-8961}, mr={2925032}}
\end{barticle}
%
\bptok{imsref}%
% NOT OUTPUTED:
% issn = 1874-8961
% url = http://dx.doi.org/10.1007/s11004-011-9377-y
% number = 4
% fjournal = Mathematical Geosciences
\endbibitem

%b28 ###
\bibitem[\protect\citeauthoryear{Du and Ma}{2013}]{DuMa13}
%
\begin{barticle}[mr]
\bauthor{\bsnm{Du},~\bfnm{Juan}\binits{J.}} \AND
\bauthor{\bsnm{Ma},~\bfnm{Chunsheng}\binits{C.}}
(\byear{2013}).
\btitle{Vector random fields with compactly supported covariance
matrix functions}.
\bjournal{J. Statist. Plann. Inference}
\bvolume{143}
\bpages{457--467}.
\bid{doi={10.1016/j.jspi.2012.08.016}, issn={0378-3758}, mr={2995107}}
\end{barticle}
%
\bptok{imsref}%
% NOT OUTPUTED:
% issn = 0378-3758
% url = http://dx.doi.org/10.1016/j.jspi.2012.08.016
% number = 3
% coden = JSPIDN
% fjournal = Journal of Statistical Planning and Inference
\endbibitem

%b29 ###
\bibitem[\protect\citeauthoryear{Du, Ma and Li}{2013}]{DuMaLi13}
%
\begin{barticle}[mr]
\bauthor{\bsnm{Du},~\bfnm{Juan}\binits{J.}},
\bauthor{\bsnm{Ma},~\bfnm{Chunsheng}\binits{C.}} \AND
\bauthor{\bsnm{Li},~\bfnm{Yang}\binits{Y.}}
(\byear{2013}).
\btitle{Isotropic variogram matrix functions on spheres}.
\bjournal{Math. Geosci.}
\bvolume{45}
\bpages{341--357}.
\bid{doi={10.1007/s11004-013-9441-x}, issn={1874-8961}, mr={3107159}}
\end{barticle}
%
\bptok{imsref}%
% NOT OUTPUTED:
% issn = 1874-8961
% url = http://dx.doi.org/10.1007/s11004-013-9441-x
% number = 3
% fjournal = Mathematical Geosciences
\endbibitem

%b30 ###
\bibitem[\protect\citeauthoryear{Du et~al.}{2012}]{Duetal12}
%
\begin{barticle}[mr]
\bauthor{\bsnm{Du},~\bfnm{Juan}\binits{J.}},
\bauthor{\bsnm{Leonenko},~\bfnm{Nikolai}\binits{N.}},
\bauthor{\bsnm{Ma},~\bfnm{Chunsheng}\binits{C.}} \AND
\bauthor{\bsnm{Shu},~\bfnm{Hong}\binits{H.}}
(\byear{2012}).
\btitle{Hyperbolic vector random fields with hyperbolic direct and
cross covariance functions}.
\bjournal{Stoch. Anal. Appl.}
\bvolume{30}
\bpages{662--674}.
\bid{doi={10.1080/07362994.2012.684325}, issn={0736-2994}, mr={2946043}}
\end{barticle}
%
\bptok{imsref}%
% NOT OUTPUTED:
% issn = 0736-2994
% url = http://dx.doi.org/10.1080/07362994.2012.684325
% number = 4
% fjournal = Stochastic Analysis and Applications
\endbibitem

%b31 ###
\bibitem[\protect\citeauthoryear{Fanshawe and Diggle}{2012}]{FanDig12}
%
\begin{barticle}[mr]
\bauthor{\bsnm{Fanshawe},~\bfnm{Thomas~R.}\binits{T.~R.}} \AND
\bauthor{\bsnm{Diggle},~\bfnm{Peter~J.}\binits{P.~J.}}
(\byear{2012}).
\btitle{Bivariate geostatistical modelling: A review and an
application to spatial variation in radon concentrations}.
\bjournal{Environ. Ecol. Stat.}
\bvolume{19}
\bpages{139--160}.
\bid{issn={1352-8505}, mr={2946168}}
\end{barticle}
%
\bptok{imsref}%
% NOT OUTPUTED:
% issn = 1352-8505
% number = 2
% coden = EESTFM
% fjournal = Environmental and Ecological Statistics
\endbibitem

%b32 ###
\bibitem[\protect\citeauthoryear{Fuentes}{2002}]{Fue02}
%
\begin{barticle}[mr]
\bauthor{\bsnm{Fuentes},~\bfnm{Montserrat}\binits{M.}}
(\byear{2002}).
\btitle{Spectral methods for nonstationary spatial processes}.
\bjournal{Biometrika}
\bvolume{89}
\bpages{197--210}.
\bid{doi={10.1093/biomet/89.1.197}, issn={0006-3444}, mr={1888368}}
\end{barticle}
%
\bptok{imsref}%
% NOT OUTPUTED:
% issn = 0006-3444
% url = http://dx.doi.org/10.1093/biomet/89.1.197
% number = 1
% coden = BIOKAX
% fjournal = Biometrika
\endbibitem

%b33 ###
\bibitem[\protect\citeauthoryear{Fuentes and Reich}{2013}]{FueRei13}
%
\begin{barticle}[mr]
\bauthor{\bsnm{Fuentes},~\bfnm{Montserrat}\binits{M.}} \AND
\bauthor{\bsnm{Reich},~\bfnm{Brian}\binits{B.}}
(\byear{2013}).
\btitle{Multivariate spatial nonparametric modelling via kernel
processes mixing}.
\bjournal{Statist. Sinica}
\bvolume{23}
\bpages{75--97}.
\bid{issn={1017-0405}, mr={3076159}}
\end{barticle}
%
\bptok{imsref}%
% NOT OUTPUTED:
% issn = 1017-0405
% number = 1
% fjournal = Statistica Sinica
\endbibitem

%b34 ###
\bibitem[\protect\citeauthoryear{Furrer}{2005}]{Fur05}
%
\begin{barticle}[mr]
\bauthor{\bsnm{Furrer},~\bfnm{Reinhard}\binits{R.}}
(\byear{2005}).
\btitle{Covariance estimation under spatial dependence}.
\bjournal{J. Multivariate Anal.}
\bvolume{94}
\bpages{366--381}.
\bid{doi={10.1016/j.jmva.2004.05.009}, issn={0047-259X}, mr={2167920}}
\end{barticle}
%
\bptok{imsref}%
% NOT OUTPUTED:
% issn = 0047-259X
% url = http://dx.doi.org/10.1016/j.jmva.2004.05.009
% number = 2
% fjournal = Journal of Multivariate Analysis
\endbibitem

%b35 ###
\bibitem[\protect\citeauthoryear{Furrer and Genton}{2011}]{FurGen11}
%
\begin{barticle}[mr]
\bauthor{\bsnm{Furrer},~\bfnm{Reinhard}\binits{R.}} \AND
\bauthor{\bsnm{Genton},~\bfnm{Marc~G.}\binits{M.~G.}}
(\byear{2011}).
\btitle{Aggregation-cokriging for highly multivariate spatial data}.
\bjournal{Biometrika}
\bvolume{98}
\bpages{615--631}.
\bid{doi={10.1093/biomet/asr029}, issn={0006-3444}, mr={2836410}}
\end{barticle}
%
\bptok{imsref}%
% NOT OUTPUTED:
% issn = 0006-3444
% url = http://dx.doi.org/10.1093/biomet/asr029
% number = 3
% coden = BIOKAX
% fjournal = Biometrika
\endbibitem

%b36 ###
\bibitem[\protect\citeauthoryear{Furrer, Genton and
Nychka}{2006}]{FurGenNyc06}
%
\begin{barticle}[mr]
\bauthor{\bsnm{Furrer},~\bfnm{Reinhard}\binits{R.}},
\bauthor{\bsnm{Genton},~\bfnm{Marc~G.}\binits{M.~G.}} \AND
\bauthor{\bsnm{Nychka},~\bfnm{Douglas}\binits{D.}}
(\byear{2006}).
\btitle{Covariance tapering for interpolation of large spatial datasets}.
\bjournal{J. Comput. Graph. Statist.}
\bvolume{15}
\bpages{502--523}.
\bid{doi={10.1198/106186006X132178}, issn={1061-8600}, mr={2291261}}
\end{barticle}
%
\bptok{imsref}%
% NOT OUTPUTED:
% issn = 1061-8600
% url = http://dx.doi.org/10.1198/106186006X132178
% number = 3
% fjournal = Journal of Computational and Graphical Statistics
\endbibitem

%b37 ###
\bibitem[\protect\citeauthoryear{Fuselier}{2008}]{Fus08}
%
\begin{barticle}[mr]
\bauthor{\bsnm{Fuselier},~\bfnm{Edward~J.}\binits{E.~J.}}
(\byear{2008}).
\btitle{Improved stability estimates and a characterization of the
native space for matrix-valued {RBF}s}.
\bjournal{Adv. Comput. Math.}
\bvolume{29}
\bpages{269--290}.
\bid{doi={10.1007/s10444-007-9046-3}, issn={1019-7168}, mr={2438345}}
\end{barticle}
%
\bptok{imsref}%
% NOT OUTPUTED:
% issn = 1019-7168
% url = http://dx.doi.org/10.1007/s10444-007-9046-3
% number = 3
% fjournal = Advances in Computational Mathematics
\endbibitem

%b38 ###
\bibitem[\protect\citeauthoryear{Gaspari and Cohn}{1999}]{GasCoh99}
%
\begin{barticle}[auto:STB|2014/06/18|12:29:53]
\bauthor{\bsnm{Gaspari},~\bfnm{G.}\binits{G.}} \AND
\bauthor{\bsnm{Cohn},~\bfnm{S.~E.}\binits{S.~E.}}
(\byear{1999}).
\btitle{Construction of correlation functions in two and three dimensions}.
\bjournal{Quarterly J. Roy. Meteorol. Soc.}
\bvolume{125}
\bpages{723--757}.
\end{barticle}
%
\bptok{imsref}%
\endbibitem

%b39 ###
\bibitem[\protect\citeauthoryear{Gaspari et~al.}{2006}]{Gasetal}
%
\begin{barticle}[auto:STB|2014/06/18|12:29:53]
\bauthor{\bsnm{Gaspari},~\bfnm{G.}\binits{G.}},
\bauthor{\bsnm{Cohn},~\bfnm{S.~E.}\binits{S.~E.}},
\bauthor{\bsnm{Guo},~\bfnm{J.}\binits{J.}} \AND
\bauthor{\bsnm{Pawson},~\bfnm{S.}\binits{S.}}
(\byear{2006}).
\btitle{Construction and application of covariance functions with
variable length-fields}.
\bjournal{Quarterly J. Roy. Meteorol. Soc.}
\bvolume{132}
\bpages{1815--1838}.
\end{barticle}
%
\bptok{imsref}%
% NOT OUTPUTED:
% sortkey = Gaspari(2006
% howpublished = (2006), ``Construction and application of covariance
%functions with variable length-fields, " \textit{Quarterly Journal of
%the Royal Meteorological Society}, 1815-1838
\endbibitem

%b40 ###
\bibitem[\protect\citeauthoryear{Gelfand, Banerjee and
Gamerman}{2005}]{GelBanGam05}
%
\begin{barticle}[mr]
\bauthor{\bsnm{Gelfand},~\bfnm{Alan~E.}\binits{A.~E.}},
\bauthor{\bsnm{Banerjee},~\bfnm{Sudipto}\binits{S.}} \AND
\bauthor{\bsnm{Gamerman},~\bfnm{Dani}\binits{D.}}
(\byear{2005}).
\btitle{Spatial process modelling for univariate and multivariate
dynamic spatial data}.
\bjournal{Environmetrics}
\bvolume{16}
\bpages{465--479}.
\bid{doi={10.1002/env.715}, issn={1180-4009}, mr={2147537}}
\end{barticle}
%
\bptok{imsref}%
% NOT OUTPUTED:
% issn = 1180-4009
% url = http://dx.doi.org/10.1002/env.715
% number = 5
% fjournal = Environmetrics
\endbibitem

%b41 ###
\bibitem[\protect\citeauthoryear{Gelfand and Vounatsou}{2003}]{GelVou03}
%
\begin{barticle}[pbm]
\bauthor{\bsnm{Gelfand},~\bfnm{Alan~E.}\binits{A.~E.}} \AND
\bauthor{\bsnm{Vounatsou},~\bfnm{Penelope}\binits{P.}}
(\byear{2003}).
\btitle{Proper multivariate conditional autoregressive models for
spatial data analysis}.
\bjournal{Biostatistics}
\bvolume{4}
\bpages{11--25}.
\bid{doi={10.1093/biostatistics/4.1.11}, issn={1465-4644},
pii={4/1/11}, pmid={12925327}}
\end{barticle}
%
\bptok{imsref}%
% NOT OUTPUTED:
% issn = 1465-4644
% number = 1
% fjournal = Biostatistics (Oxford, England)
\endbibitem

%b42 ###
\bibitem[\protect\citeauthoryear{Gelfand et~al.}{2004}]{Geletal04}
%
\begin{barticle}[mr]
\bauthor{\bsnm{Gelfand},~\bfnm{Alan~E.}\binits{A.~E.}},
\bauthor{\bsnm{Schmidt},~\bfnm{Alexandra~M.}\binits{A.~M.}},
\bauthor{\bsnm{Banerjee},~\bfnm{Sudipto}\binits{S.}} \AND
\bauthor{\bsnm{Sirmans},~\bfnm{C.~F.}\binits{C.~F.}}
(\byear{2004}).
\btitle{Nonstationary multivariate process modeling through spatially
varying coregionalization}.
\bjournal{Test}
\bvolume{13}
\bpages{263--312}.
%\bnote{With discussion by Montserrat Fuentes, Dave Higdon and Bruno
%Sans{\'o} and a rejoinder by the authors}.
\bid{doi={10.1007/BF02595775}, issn={1133-0686}, mr={2154003}}
\bptnote{check related}%
\end{barticle}
%
\bptok{imsref}%
% NOT OUTPUTED:
% issn = 1133-0686
% url = http://dx.doi.org/10.1007/BF02595775
% number = 2
% fjournal = Test
\endbibitem

%b43 ###
\bibitem[\protect\citeauthoryear{Genton, Padoan and Sang}{2015}]{GenPadSan}
%
\begin{barticle}[auto:STB|2014/06/18|12:29:53]
\bauthor{\bsnm{Genton},~\bfnm{M.~G.}\binits{M.~G.}},
\bauthor{\bsnm{Padoan},~\bfnm{S.}\binits{S.}} \AND
\bauthor{\bsnm{Sang},~\bfnm{H.}\binits{H.}}
(\byear{2015}).
\btitle{Multivariate max-stable spatial processes}.
\bjournal{Biometrika}
\bvolume{102}
\bpages{215--230}.
\bid{mr={3335107}}
\end{barticle}
%
\bptok{imsref}%
% NOT OUTPUTED:
% sortkey = Genton(2014
% howpublished = (2014), ``Multivariate max-stable spatial processes,
%'' manuscript
\endbibitem

%b44 ###
\bibitem[\protect\citeauthoryear{Gneiting}{2002}]{Gne02}
%
\begin{barticle}[mr]
\bauthor{\bsnm{Gneiting},~\bfnm{Tilmann}\binits{T.}}
(\byear{2002}).
\btitle{Nonseparable, stationary covariance functions for space--time data}.
\bjournal{J. Amer. Statist. Assoc.}
\bvolume{97}
\bpages{590--600}.
\bid{doi={10.1198/016214502760047113}, issn={0162-1459}, mr={1941475}}
\end{barticle}
%
\bptok{imsref}%
% NOT OUTPUTED:
% issn = 0162-1459
% url = http://dx.doi.org/10.1198/016214502760047113
% number = 458
% coden = JSTNAL
% fjournal = Journal of the American Statistical Association
\endbibitem

%b45 ###
\bibitem[\protect\citeauthoryear{Gneiting}{2013}]{Gne13}
%
\begin{barticle}[mr]
\bauthor{\bsnm{Gneiting},~\bfnm{Tilmann}\binits{T.}}
(\byear{2013}).
\btitle{Strictly and non-strictly positive definite functions on spheres}.
\bjournal{Bernoulli}
\bvolume{19}
\bpages{1327--1349}.
\bid{doi={10.3150/12-BEJSP06}, issn={1350-7265}, mr={3102554}}
\end{barticle}
%
\bptok{imsref}%
% NOT OUTPUTED:
% issn = 1350-7265
% url = http://dx.doi.org/10.3150/12-BEJSP06
% number = 4
% fjournal = Bernoulli. Official Journal of the Bernoulli Society for
%Mathematical Statistics and Probability
\endbibitem

%b46 ###
\bibitem[\protect\citeauthoryear{Gneiting, Genton and Guttorp}{2007}]{autokey46}
\begin{bincollection}[auto:STB|2014/06/18|12:29:53]
\bauthor{\bsnm{Gneiting},~\bfnm{Tilmann}\binits{T.}},
\bauthor{\bsnm{Genton},~\bfnm{M. G.}\binits{M. G.}} \AND
\bauthor{\bsnm{Guttorp},~\bfnm{P.}\binits{P.}}
(\byear{2007}).
\btitle{Geostatistical space-time models, stationarity, separability and full
symmetry}.
In \bbooktitle{Statistics of Spatio-Temporal Systems. Monographs in
Statistics and Applied Probability}
(\beditor{\binits{B.}~\bsnm{Finkenstaedt}},
\beditor{\binits{L.}~\bsnm{Held}} \AND
\beditor{\binits{V.}~\bsnm{Isham}}, eds.)
\bpages{151--175}.
\bpublisher{Chapman \& Hall/CRC Press},
\blocation{Boca Raton, FL}.
\end{bincollection}
%
\bptok{imsref}%
\endbibitem

%b47 ###
\bibitem[\protect\citeauthoryear{Gneiting and Guttorp}{2006}]{GutGne06}
%
\begin{barticle}[mr]
\bauthor{\bsnm{Gneiting},~\bfnm{Tilmann}\binits{T.}} \AND
\bauthor{\bsnm{Guttorp},~\bfnm{Peter}\binits{P.}}
(\byear{2006}).
\btitle{Studies in the history of probability and statistics. XLIX.
{O}n the {M}at\'ern correlation family}.
\bjournal{Biometrika}
\bvolume{93}
\bpages{989--995}.
\bid{doi={10.1093/biomet/93.4.989}, issn={0006-3444}, mr={2285084}}
\end{barticle}
%
\bptok{imsref}%
% NOT OUTPUTED:
% issn = 0006-3444
% url = http://dx.doi.org/10.1093/biomet/93.4.989
% number = 4
% coden = BIOKAX
% fjournal = Biometrika
\endbibitem

%b48 ###
\bibitem[\protect\citeauthoryear{Gneiting, Kleiber and
Schlather}{2010}]{GneKleSch10}
%
\begin{barticle}[mr]
\bauthor{\bsnm{Gneiting},~\bfnm{Tilmann}\binits{T.}},
\bauthor{\bsnm{Kleiber},~\bfnm{William}\binits{W.}} \AND
\bauthor{\bsnm{Schlather},~\bfnm{Martin}\binits{M.}}
(\byear{2010}).
\btitle{Mat\'ern cross-covariance functions for multivariate random fields}.
\bjournal{J. Amer. Statist. Assoc.}
\bvolume{105}
\bpages{1167--1177}.
\bid{doi={10.1198/jasa.2010.tm09420}, issn={0162-1459}, mr={2752612}}
\end{barticle}
%
\bptok{imsref}%
% NOT OUTPUTED:
% issn = 0162-1459
% url = http://dx.doi.org/10.1198/jasa.2010.tm09420
% number = 491
% coden = JSTNAL
% fjournal = Journal of the American Statistical Association
\endbibitem

%b49 ###
\bibitem[\protect\citeauthoryear{Gneiting and Schlather}{2004}]{GneSch04}
%
\begin{barticle}[mr]
\bauthor{\bsnm{Gneiting},~\bfnm{Tilmann}\binits{T.}} \AND
\bauthor{\bsnm{Schlather},~\bfnm{Martin}\binits{M.}}
(\byear{2004}).
\btitle{Stochastic models that separate fractal dimension and the
{H}urst effect}.
\bjournal{SIAM Rev.}
\bvolume{46}
\bpages{269--282 (electronic)}.
\bid{doi={10.1137/S0036144501394387}, issn={0036-1445}, mr={2114455}}
\end{barticle}
%
\bptok{imsref}%
% NOT OUTPUTED:
% issn = 0036-1445
% url = http://dx.doi.org/10.1137/S0036144501394387
% number = 2
% fjournal = SIAM Review
\endbibitem

%b50 ###
\bibitem[\protect\citeauthoryear{Goff and Jordan}{1988}]{GofJor88}
%
\begin{barticle}[auto:STB|2014/06/18|12:29:53]
\bauthor{\bsnm{Goff},~\bfnm{J.~A.}\binits{J.~A.}} \AND
\bauthor{\bsnm{Jordan},~\bfnm{T.~H.}\binits{T.~H.}}
(\byear{1988}).
\btitle{Stochastic modeling of seafloor morphology: inversion of sea
beam data for second-order statistics}.
\bjournal{J. Geophys. Res.}
\bvolume{93}
\bpages{13589--13608}.
\end{barticle}
%
\bptok{imsref}%
\endbibitem

%b51 ###
\bibitem[\protect\citeauthoryear{Goulard and Voltz}{1992}]{GouVol92}
%
\begin{barticle}[auto:STB|2014/06/18|12:29:53]
\bauthor{\bsnm{Goulard},~\bfnm{M.}\binits{M.}} \AND
\bauthor{\bsnm{Voltz},~\bfnm{M.}\binits{M.}}
(\byear{1992}).
\btitle{Linear coregionalization model: Tools for estimation and
choice of cross-variogram matrix}.
\bjournal{Math. Geol.}
\bvolume{24}
\bpages{269--282}.
\end{barticle}
%
\bptok{imsref}%
\endbibitem

%b52 ###
\bibitem[\protect\citeauthoryear{Grzebyk and Wackernagel}{1994}]{GrzWac94}
%
\begin{bincollection}[auto:STB|2014/06/18|12:29:53]
\bauthor{\bsnm{Grzebyk},~\bfnm{M.}\binits{M.}} \AND
\bauthor{\bsnm{Wackernagel},~\bfnm{H.}\binits{H.}}
(\byear{1994}).
\btitle{Multivariate analysis and spatial/temporal scales: Real and
complex models}.
In \bbooktitle{Proceedings of XVIIth International Biometric
Conference, Hamilton, Ontario, Canada}
\bvolume{1}
\bpages{19--33}.
\end{bincollection}
%
\bptok{imsref}%
\endbibitem

%b53 ###
\bibitem[\protect\citeauthoryear{Guhaniyogi et~al.}{2013}]{Guhetal13}
%
\begin{barticle}[mr]
\bauthor{\bsnm{Guhaniyogi},~\bfnm{Rajarshi}\binits{R.}},
\bauthor{\bsnm{Finley},~\bfnm{Andrew~O.}\binits{A.~O.}},
\bauthor{\bsnm{Banerjee},~\bfnm{Sudipto}\binits{S.}} \AND
\bauthor{\bsnm{Kobe},~\bfnm{Richard~K.}\binits{R.~K.}}
(\byear{2013}).
\btitle{Modeling complex spatial dependencies: Low-rank spatially
varying cross-covariances with application to soil nutrient data}.
\bjournal{J. Agric. Biol. Environ. Stat.}
\bvolume{18}
\bpages{274--298}.
\bid{doi={10.1007/s13253-013-0140-3}, issn={1085-7117}, mr={3110894}}
\end{barticle}
%
\bptok{imsref}%
% NOT OUTPUTED:
% issn = 1085-7117
% url = http://dx.doi.org/10.1007/s13253-013-0140-3
% number = 3
% fjournal = Journal of Agricultural, Biological, and Environmental
%Statistics
\endbibitem

%b54 ###
\bibitem[\protect\citeauthoryear{Guillot, Senoussi and
Monestiez}{2001}]{GuiSenMon01}
%
\begin{bincollection}[auto:STB|2014/06/18|12:29:53]
\bauthor{\bsnm{Guillot},~\bfnm{G.}\binits{G.}},
\bauthor{\bsnm{Senoussi},~\bfnm{R.}\binits{R.}} \AND
\bauthor{\bsnm{Monestiez},~\bfnm{P.}\binits{P.}}
(\byear{2001}).
\btitle{A~positive definite estimator of the non stationary covariance
of random fields}.
In \bbooktitle{GeoENV 2000: Third European Conference on Geostatistics
for Environmental Applications}
(\beditor{\bfnm{P.}\binits{P.}~\bsnm{Monestiez}},
\beditor{\bfnm{D.}\binits{D.}~\bsnm{Allard}} \AND
\beditor{\bfnm{R.}\binits{R.}~\bsnm{Froidevaux}}, eds.)
\bpages{333--344}.
\bpublisher{Kluwer Academic},
\blocation{Dordrecht}.
\end{bincollection}
%
\bptok{imsref}%
\endbibitem

%b55 ###
\bibitem[\protect\citeauthoryear{Helterbrand and Cressie}{1994}]{HelCre94}
%
\begin{barticle}[mr]
\bauthor{\bsnm{Helterbrand},~\bfnm{Jeffrey~D.}\binits{J.~D.}} \AND
\bauthor{\bsnm{Cressie},~\bfnm{Noel}\binits{N.}}
(\byear{1994}).
\btitle{Universal co-{k}riging under intrinsic coregionalization}.
\bjournal{Math. Geol.}
\bvolume{26}
\bpages{205--226}.
\bid{doi={10.1007/BF02082764}, issn={0882-8121}, mr={1267008}}
\end{barticle}
%
\bptok{imsref}%
% NOT OUTPUTED:
% issn = 0882-8121
% url = http://dx.doi.org/10.1007/BF02082764
% number = 2
% coden = MATGED
% fjournal = Mathematical Geology
\endbibitem

%b56 ###
\bibitem[\protect\citeauthoryear{Hristopoulos and Porcu}{2014}]{HriPor}
%
\begin{barticle}[auto:STB|2014/06/18|12:29:53]
\bauthor{\bsnm{Hristopoulos},~\bfnm{D.}\binits{D.}} \AND
\bauthor{\bsnm{Porcu},~\bfnm{E.}\binits{E.}}
(\byear{2014}).
\btitle{Vector Spartan spatial random field models}.
\bjournal{Probabilistic Engineering Mechanics}
\bvolume{37}
\bpages{84--92}.
\end{barticle}
%
\bptok{imsref}%
% NOT OUTPUTED:
% sortkey = Hristopoulos(2013
% howpublished = (2013), ``Vector Spartan spatial random field models,
%'' \textit{Technical Report University Federico Santa Maria}
\endbibitem

%b57 ###
\bibitem[\protect\citeauthoryear{Hu et~al.}{2013}]{Huetal}
%
\begin{bmisc}[auto:STB|2014/06/18|12:29:53]
\bauthor{\bsnm{Hu},~\bfnm{X.}\binits{X.}},
\bauthor{\bsnm{Steinsland},~\bfnm{I.}\binits{I.}},
\bauthor{\bsnm{Simpson},~\bfnm{D.}\binits{D.}},
\bauthor{\bsnm{Martino},~\bfnm{S.}\binits{S.}} \AND
\bauthor{\bsnm{Rue},~\bfnm{H.}\binits{H.}}
(\byear{2013}).
\bhowpublished{Spatial modelling of temperature and humidity using
systems of stochastic partial
differential equations. Available at \arxivurl{arXiv:1307.1402v1}}.
\end{bmisc}
%
\bptok{imsref}%
% NOT OUTPUTED:
% sortkey = Hu(2013
% howpublished = (2013), ``Spatial modelling of temperature and
%humidity using systems of stochastic partial differential equations,
%'' arXiv:1307.1402v1
\endbibitem

%b58 ###
\bibitem[\protect\citeauthoryear{Huang, Yao, Cressie and Hsing}{2009}]{Huaetal09}
%
\begin{barticle}[mr]
\bauthor{\bsnm{Huang},~\bfnm{Chunfeng}\binits{C.}},
\bauthor{\bsnm{Yao},~\bfnm{Yonggang}\binits{Y.}},
\bauthor{\bsnm{Cressie},~\bfnm{Noel}\binits{N.}} \AND
\bauthor{\bsnm{Hsing},~\bfnm{Tailen}\binits{T.}}
(\byear{2009}).
\btitle{Multivariate intrinsic random functions for cokriging}.
\bjournal{Math. Geosci.}
\bvolume{41}
\bpages{887--904}.
\bid{doi={10.1007/s11004-009-9218-4}, issn={1874-8961}, mr={2557237}}
\end{barticle}
%
\bptok{imsref}%
% NOT OUTPUTED:
% issn = 1874-8961
% url = http://dx.doi.org/10.1007/s11004-009-9218-4
% number = 8
% fjournal = Mathematical Geosciences
\endbibitem

%b59 ###
\bibitem[\protect\citeauthoryear{Journel}{1999}]{Jou99}
%
\begin{barticle}[mr]
\bauthor{\bsnm{Journel},~\bfnm{A.~G.}\binits{A.~G.}}
(\byear{1999}).
\btitle{Markov models for cross-covariances}.
\bjournal{Math. Geol.}
\bvolume{31}
\bpages{955--964}.
\bid{doi={10.1023/A:1007553013388}, issn={0882-8121}, mr={1717033}}
\end{barticle}
%
\bptok{imsref}%
% NOT OUTPUTED:
% issn = 0882-8121
% url = http://dx.doi.org/10.1023/A:1007553013388
% number = 8
% coden = MATGED
% fjournal = Mathematical Geology
\endbibitem

%b60 ###
\bibitem[\protect\citeauthoryear{Jun}{2011}]{Jun11}
%
\begin{barticle}[mr]
\bauthor{\bsnm{Jun},~\bfnm{Mikyoung}\binits{M.}}
(\byear{2011}).
\btitle{Non-stationary cross-covariance models for multivariate
processes on a globe}.
\bjournal{Scand. J. Stat.}
\bvolume{38}
\bpages{726--747}.
\bid{doi={10.1111/j.1467-9469.2011.00751.x}, issn={0303-6898}, mr={2859747}}
\end{barticle}
%
\bptok{imsref}%
% NOT OUTPUTED:
% issn = 0303-6898
% url = http://dx.doi.org/10.1111/j.1467-9469.2011.00751.x
% number = 4
% fjournal = Scandinavian Journal of Statistics. Theory and Applications
\endbibitem

%b61 ###
\bibitem[\protect\citeauthoryear{Jun}{2014}]{Jun14}
%
\begin{barticle}[mr]
\bauthor{\bsnm{Jun},~\bfnm{Mikyoung}\binits{M.}}
(\byear{2014}).
\btitle{Mat\'ern-based nonstationary cross-covariance models for
global processes}.
\bjournal{J. Multivariate Anal.}
\bvolume{128}
\bpages{134--146}.
\bid{doi={10.1016/j.jmva.2014.03.009}, issn={0047-259X}, mr={3199833}}
\end{barticle}
%
\bptok{imsref}%
% NOT OUTPUTED:
% issn = 0047-259X
% url = http://dx.doi.org/10.1016/j.jmva.2014.03.009
% fjournal = Journal of Multivariate Analysis
\endbibitem

%b62 ###
\bibitem[\protect\citeauthoryear{Jun and Genton}{2012}]{JunGen12}
%
\begin{barticle}[mr]
\bauthor{\bsnm{Jun},~\bfnm{Mikyoung}\binits{M.}} \AND
\bauthor{\bsnm{Genton},~\bfnm{Marc~G.}\binits{M.~G.}}
(\byear{2012}).
\btitle{A test for stationarity of spatio-temporal random fields on
planar and spherical domains}.
\bjournal{Statist. Sinica}
\bvolume{22}
\bpages{1737--1764}.
\bid{issn={1017-0405}, mr={3027105}}
\end{barticle}
%
\bptok{imsref}%
% NOT OUTPUTED:
% issn = 1017-0405
% number = 4
% fjournal = Statistica Sinica
\endbibitem

%b63 ###
\bibitem[\protect\citeauthoryear{Jun et~al.}{2011}]{Junetal11}
%
\begin{barticle}[auto:STB|2014/06/18|12:29:53]
\bauthor{\bsnm{Jun},~\bfnm{M.}\binits{M.}},
\bauthor{\bsnm{Szunyogh},~\bfnm{I.}\binits{I.}},
\bauthor{\bsnm{Genton},~\bfnm{M.~G.}\binits{M.~G.}},
\bauthor{\bsnm{Zhang},~\bfnm{F.}\binits{F.}} \AND
\bauthor{\bsnm{Bishop},~\bfnm{C.~H.}\binits{C.~H.}}
(\byear{2011}).
\btitle{A statistical investigation of the sensitivity of
ensemble-based Kalman filters to covariance filtering}.
\bjournal{Mon. Weather Rev.}
\bvolume{139}
\bpages{3036--3051}.
\end{barticle}
%
\bptok{imsref}%
\endbibitem

%b64 ###
\bibitem[\protect\citeauthoryear{Kaufman, Schervish and
Nychka}{2008}]{KauSchNyc08}
%
\begin{barticle}[mr]
\bauthor{\bsnm{Kaufman},~\bfnm{Cari~G.}\binits{C.~G.}},
\bauthor{\bsnm{Schervish},~\bfnm{Mark~J.}\binits{M.~J.}} \AND
\bauthor{\bsnm{Nychka},~\bfnm{Douglas~W.}\binits{D.~W.}}
(\byear{2008}).
\btitle{Covariance tapering for likelihood-based estimation in large
spatial data sets}.
\bjournal{J. Amer. Statist. Assoc.}
\bvolume{103}
\bpages{1545--1555}.
\bid{doi={10.1198/016214508000000959}, issn={0162-1459}, mr={2504203}}
\end{barticle}
%
\bptok{imsref}%
% NOT OUTPUTED:
% issn = 0162-1459
% url = http://dx.doi.org/10.1198/016214508000000959
% number = 484
% coden = JSTNAL
% fjournal = Journal of the American Statistical Association
\endbibitem

%b65 ###
\bibitem[\protect\citeauthoryear{Kleiber and Genton}{2013}]{KleGen13}
%
\begin{barticle}[mr]
\bauthor{\bsnm{Kleiber},~\bfnm{William}\binits{W.}} \AND
\bauthor{\bsnm{Genton},~\bfnm{Marc~G.}\binits{M.~G.}}
(\byear{2013}).
\btitle{Spatially varying cross-correlation coefficients in the
presence of nugget effects}.
\bjournal{Biometrika}
\bvolume{100}
\bpages{213--220}.
\bid{doi={10.1093/biomet/ass057}, issn={0006-3444}, mr={3034334}}
\end{barticle}
%
\bptok{imsref}%
% NOT OUTPUTED:
% issn = 0006-3444
% url = http://dx.doi.org/10.1093/biomet/ass057
% number = 1
% fjournal = Biometrika
\endbibitem

%b66 ###
\bibitem[\protect\citeauthoryear{Kleiber, Katz and
Rajagopalan}{2013}]{KleKatRaj13}
%
\begin{barticle}[mr]
\bauthor{\bsnm{Kleiber},~\bfnm{William}\binits{W.}},
\bauthor{\bsnm{Katz},~\bfnm{Richard~W.}\binits{R.~W.}} \AND
\bauthor{\bsnm{Rajagopalan},~\bfnm{Balaji}\binits{B.}}
(\byear{2013}).
\btitle{Daily minimum and maximum temperature simulation over complex terrain}.
\bjournal{Ann. Appl. Stat.}
\bvolume{7}
\bpages{588--612}.
\bid{doi={10.1214/12-AOAS602}, issn={1932-6157}, mr={3086432}}
\end{barticle}
%
\bptok{imsref}%
% NOT OUTPUTED:
% issn = 1932-6157
% url = http://dx.doi.org/10.1214/12-AOAS602
% number = 1
% fjournal = The Annals of Applied Statistics
\endbibitem

%b67 ###
\bibitem[\protect\citeauthoryear{Kleiber and Nychka}{2012}]{KleNyc12}
%
\begin{barticle}[mr]
\bauthor{\bsnm{Kleiber},~\bfnm{William}\binits{W.}} \AND
\bauthor{\bsnm{Nychka},~\bfnm{Douglas}\binits{D.}}
(\byear{2012}).
\btitle{Nonstationary modeling for multivariate spatial processes}.
\bjournal{J. Multivariate Anal.}
\bvolume{112}
\bpages{76--91}.
\bid{doi={10.1016/j.jmva.2012.05.011}, issn={0047-259X}, mr={2957287}}
\end{barticle}
%
\bptok{imsref}%
% NOT OUTPUTED:
% issn = 0047-259X
% url = http://dx.doi.org/10.1016/j.jmva.2012.05.011
% fjournal = Journal of Multivariate Analysis
\endbibitem

%b68 ###
\bibitem[\protect\citeauthoryear{Kleiber and Porcu}{2015}]{KlePor}
%
\begin{barticle}[auto:STB|2014/06/18|12:29:53]
\bauthor{\bsnm{Kleiber},~\bfnm{W.}\binits{W.}} \AND
\bauthor{\bsnm{Porcu},~\bfnm{E.}\binits{E.}}
(\byear{2015}).
\btitle{Nonstationary matrix covariances: Compact support, long
range dependence and quasi-arithmetic constructions}.
\bjournal{Stoch. Environ. Res. Risk Assess.}
\bvolume{29}
\bpages{193--204}.
\end{barticle}
%
\bptok{imsref}%
% NOT OUTPUTED:
% sortkey = Kleiber(2014
% howpublished = (2014), ``Nonstationary matrix covariances: compact
%support, long range dependence and quasi-arithmetic constructions, ''
%\textit{Stochastic Environmental Research and Risk Assessment}, in
%press
\endbibitem

%b69 ###
\bibitem[\protect\citeauthoryear{K{\"u}nsch, Papritz and
Bassi}{1997}]{KunPapBas97}
%
\begin{barticle}[mr]
\bauthor{\bsnm{K{\"u}nsch},~\bfnm{H.~R.}\binits{H.~R.}},
\bauthor{\bsnm{Papritz},~\bfnm{A.}\binits{A.}} \AND
\bauthor{\bsnm{Bassi},~\bfnm{F.}\binits{F.}}
(\byear{1997}).
\btitle{Generalized cross-covariances and their estimation}.
\bjournal{Math. Geol.}
\bvolume{29}
\bpages{779--799}.
\bid{doi={10.1007/BF02768902}, issn={0882-8121}, mr={1470669}}
\end{barticle}
%
\bptok{imsref}%
% NOT OUTPUTED:
% issn = 0882-8121
% url = http://dx.doi.org/10.1007/BF02768902
% number = 6
% coden = MATGED
% fjournal = Mathematical Geology
\endbibitem

%b70 ###
\bibitem[\protect\citeauthoryear{Lark}{2003}]{Lar03}
%
\begin{barticle}[auto:STB|2014/06/18|12:29:53]
\bauthor{\bsnm{Lark},~\bfnm{R.~M.}\binits{R.~M.}}
(\byear{2003}).
\btitle{Two robust estimators of the cross-variogram for multivariate
geostatistical analysis of soil properties}.
\bjournal{European J. Soil Sci.}
\bvolume{54}
\bpages{187--201}.
\end{barticle}
%
\bptok{imsref}%
\endbibitem

%b71 ###
\bibitem[\protect\citeauthoryear{Li, Genton and Sherman}{2008}]{LiGenShe08}
%
\begin{barticle}[mr]
\bauthor{\bsnm{Li},~\bfnm{Bo}\binits{B.}},
\bauthor{\bsnm{Genton},~\bfnm{Marc~G.}\binits{M.~G.}} \AND
\bauthor{\bsnm{Sherman},~\bfnm{Michael}\binits{M.}}
(\byear{2008}).
\btitle{Testing the covariance structure of multivariate random fields}.
\bjournal{Biometrika}
\bvolume{95}
\bpages{813--829}.
\bid{doi={10.1093/biomet/asn053}, issn={0006-3444}, mr={2461213}}
\end{barticle}
%
\bptok{imsref}%
% NOT OUTPUTED:
% issn = 0006-3444
% url = http://dx.doi.org/10.1093/biomet/asn053
% number = 4
% coden = BIOKAX
% fjournal = Biometrika
\endbibitem

%b72 ###
\bibitem[\protect\citeauthoryear{Li and Zhang}{2011}]{LiZha11}
%
\begin{barticle}[mr]
\bauthor{\bsnm{Li},~\bfnm{Bo}\binits{B.}} \AND
\bauthor{\bsnm{Zhang},~\bfnm{Hao}\binits{H.}}
(\byear{2011}).
\btitle{An approach to modeling asymmetric multivariate spatial
covariance structures}.
\bjournal{J. Multivariate Anal.}
\bvolume{102}
\bpages{1445--1453}.
\bid{doi={10.1016/j.jmva.2011.05.010}, issn={0047-259X}, mr={2819961}}
\end{barticle}
%
\bptok{imsref}%
% NOT OUTPUTED:
% issn = 0047-259X
% url = http://dx.doi.org/10.1016/j.jmva.2011.05.010
% number = 10
% fjournal = Journal of Multivariate Analysis
\endbibitem

%b73 ###
\bibitem[\protect\citeauthoryear{Lim and Stein}{2008}]{LimSte08}
%
\begin{barticle}[mr]
\bauthor{\bsnm{Lim},~\bfnm{Chae~Young}\binits{C.~Y.}} \AND
\bauthor{\bsnm{Stein},~\bfnm{Michael}\binits{M.}}
(\byear{2008}).
\btitle{Properties of spatial cross-periodograms using fixed-domain
asymptotics}.
\bjournal{J. Multivariate Anal.}
\bvolume{99}
\bpages{1962--1984}.
\bid{doi={10.1016/j.jmva.2008.02.005}, issn={0047-259X}, mr={2466546}}
\end{barticle}
%
\bptok{imsref}%
% NOT OUTPUTED:
% issn = 0047-259X
% url = http://dx.doi.org/10.1016/j.jmva.2008.02.005
% number = 9
% fjournal = Journal of Multivariate Analysis
\endbibitem

%b74 ###
\bibitem[\protect\citeauthoryear{Long and Myers}{1997}]{LonMye97}
%
\begin{barticle}[mr]
\bauthor{\bsnm{Long},~\bfnm{Andrew~E.}\binits{A.~E.}} \AND
\bauthor{\bsnm{Myers},~\bfnm{Donald~E.}\binits{D.~E.}}
(\byear{1997}).
\btitle{A new form of the cokriging equations}.
\bjournal{Math. Geol.}
\bvolume{29}
\bpages{685--703}.
\bid{doi={10.1007/BF02769651}, issn={0882-8121}, mr={1460688}}
\end{barticle}
%
\bptok{imsref}%
% NOT OUTPUTED:
% issn = 0882-8121
% url = http://dx.doi.org/10.1007/BF02769651
% number = 5
% coden = MATGED
% fjournal = Mathematical Geology
\endbibitem

%b75 ###
\bibitem[\protect\citeauthoryear{Ma}{2011a}]{Ma11N1}
%
\begin{barticle}[mr]
\bauthor{\bsnm{Ma},~\bfnm{Chunsheng}\binits{C.}}
(\byear{2011}a).
\btitle{Vector random fields with second-order moments or second-order
increments}.
\bjournal{Stoch. Anal. Appl.}
\bvolume{29}
\bpages{197--215}.
\bid{doi={10.1080/07362994.2011.532039}, issn={0736-2994}, mr={2774237}}
\end{barticle}
%
\bptok{imsref}%
% NOT OUTPUTED:
% issn = 0736-2994
% url = http://dx.doi.org/10.1080/07362994.2011.532039
% number = 2
% fjournal = Stochastic Analysis and Applications
\endbibitem

%b76 ###
\bibitem[\protect\citeauthoryear{Ma}{2011b}]{Ma11N2}
%
\begin{barticle}[mr]
\bauthor{\bsnm{Ma},~\bfnm{Chunsheng}\binits{C.}}
(\byear{2011}b).
\btitle{Covariance matrices for second-order vector random fields in
space and time}.
\bjournal{IEEE Trans. Signal Process.}
\bvolume{59}
\bpages{2160--2168}.
\bid{doi={10.1109/TSP.2011.2112651}, issn={1053-587X}, mr={2816489}}
\end{barticle}
%
\bptok{imsref}%
% NOT OUTPUTED:
% issn = 1053-587X
% url = http://dx.doi.org/10.1109/TSP.2011.2112651
% number = 5
% coden = ITPRED
% fjournal = IEEE Transactions on Signal Processing
\endbibitem

%b77 ###
\bibitem[\protect\citeauthoryear{Ma}{2011c}]{Ma11N3}
%
\begin{barticle}[mr]
\bauthor{\bsnm{Ma},~\bfnm{Chunsheng}\binits{C.}}
(\byear{2011}c).
\btitle{Vector random fields with long-range dependence}.
\bjournal{Fractals}
\bvolume{19}
\bpages{249--258}.
\bid{doi={10.1142/S0218348X11005312}, issn={0218-348X}, mr={2803242}}
\end{barticle}
%
\bptok{imsref}%
% NOT OUTPUTED:
% issn = 0218-348X
% url = http://dx.doi.org/10.1142/S0218348X11005312
% number = 2
% fjournal = Fractals. Complex Geometry, Patterns, and Scaling in
%Nature and Society
\endbibitem

%b78 ###
\bibitem[\protect\citeauthoryear{Ma}{2011d}]{Ma11N4}
%
\begin{barticle}[mr]
\bauthor{\bsnm{Ma},~\bfnm{Chunsheng}\binits{C.}}
(\byear{2011}d).
\btitle{A class of variogram matrices for vector random fields in
space and/or time}.
\bjournal{Math. Geosci.}
\bvolume{43}
\bpages{229--242}.
\bid{doi={10.1007/s11004-010-9310-9}, issn={1874-8961}, mr={2764536}}
\end{barticle}
%
\bptok{imsref}%
% NOT OUTPUTED:
% issn = 1874-8961
% url = http://dx.doi.org/10.1007/s11004-010-9310-9
% number = 2
% fjournal = Mathematical Geosciences
\endbibitem

%b79 ###
\bibitem[\protect\citeauthoryear{Ma}{2012}]{Ma12}
%
\begin{barticle}[mr]
\bauthor{\bsnm{Ma},~\bfnm{Chunsheng}\binits{C.}}
(\byear{2012}).
\btitle{Stationary and isotropic vector random fields on spheres}.
\bjournal{Math. Geosci.}
\bvolume{44}
\bpages{765--778}.
\bid{doi={10.1007/s11004-012-9411-8}, issn={1874-8961}, mr={2956272}}
\end{barticle}
%
\bptok{imsref}%
% NOT OUTPUTED:
% issn = 1874-8961
% url = http://dx.doi.org/10.1007/s11004-012-9411-8
% number = 6
% fjournal = Mathematical Geosciences
\endbibitem

%b80 ###
\bibitem[\protect\citeauthoryear{Ma}{2013a}]{Ma13}
%
\begin{barticle}[mr]
\bauthor{\bsnm{Ma},~\bfnm{Chunsheng}\binits{C.}}
(\byear{2013}a).
\btitle{Student's {$t$} vector random fields with power-law and
log-law decaying direct and cross covariances}.
\bjournal{Stoch. Anal. Appl.}
\bvolume{31}
\bpages{167--182}.
\bid{doi={10.1080/07362994.2013.741401}, issn={0736-2994}, mr={3007889}}
\end{barticle}
%
\bptok{imsref}%
% NOT OUTPUTED:
% issn = 0736-2994
% url = http://dx.doi.org/10.1080/07362994.2013.741401
% number = 1
% fjournal = Stochastic Analysis and Applications
\endbibitem

%b81 ###
\bibitem[\protect\citeauthoryear{Ma}{2013b}]{Ma}
%
\begin{barticle}[auto:STB|2014/06/18|12:29:53]
\bauthor{\bsnm{Ma},~\bfnm{C.}\binits{C.}}
(\byear{2013}b).
\btitle{Mittag--Leffler vector random fields with Mittag--Leffler
direct and cross covariance functions}.
\bjournal{Ann. Inst. Statist. Math.}
\bvolume{65}
\bpages{941--958}.
\bid{mr={3105803}}
\end{barticle}
%
\bptok{imsref}%
% NOT OUTPUTED:
% sortkey = Ma
% howpublished = (2013b), ``Mittag-Leffler vector random fields with
%Mittag-Leffler direct and cross covariance functions, " \textit{Annals
%of the Institute of Statistical Mathematics}, in press
\endbibitem

%b82 ###
\bibitem[\protect\citeauthoryear{Majumdar and Gelfand}{2007}]{MajGel07}
%
\begin{barticle}[mr]
\bauthor{\bsnm{Majumdar},~\bfnm{Anandamayee}\binits{A.}} \AND
\bauthor{\bsnm{Gelfand},~\bfnm{Alan~E.}\binits{A.~E.}}
(\byear{2007}).
\btitle{Multivariate spatial modeling for geostatistical data using
convolved covariance functions}.
\bjournal{Math. Geol.}
\bvolume{39}
\bpages{225--245}.
\bid{doi={10.1007/s11004-006-9072-6}, issn={0882-8121}, mr={2324633}}
\end{barticle}
%
\bptok{imsref}%
% NOT OUTPUTED:
% issn = 0882-8121
% url = http://dx.doi.org/10.1007/s11004-006-9072-6
% number = 2
% fjournal = Mathematical Geology
\endbibitem

%b83 ###
\bibitem[\protect\citeauthoryear{Majumdar, Paul and
Bautista}{2010}]{MajPauBau10}
%
\begin{barticle}[mr]
\bauthor{\bsnm{Majumdar},~\bfnm{Anandamayee}\binits{A.}},
\bauthor{\bsnm{Paul},~\bfnm{Debashis}\binits{D.}} \AND
\bauthor{\bsnm{Bautista},~\bfnm{Dianne}\binits{D.}}
(\byear{2010}).
\btitle{A generalized convolution model for multivariate nonstationary
spatial processes}.
\bjournal{Statist. Sinica}
\bvolume{20}
\bpages{675--695}.
\bid{issn={1017-0405}, mr={2682636}}
\end{barticle}
%
\bptok{imsref}%
% NOT OUTPUTED:
% issn = 1017-0405
% number = 2
% fjournal = Statistica Sinica
\endbibitem

%b84 ###
\bibitem[\protect\citeauthoryear{Mardia and Goodall}{1993}]{MarGoo93}
%
\begin{bincollection}[mr]
\bauthor{\bsnm{Mardia},~\bfnm{Kanti~V.}\binits{K.~V.}} \AND
\bauthor{\bsnm{Goodall},~\bfnm{Colin~R.}\binits{C.~R.}}
(\byear{1993}).
\btitle{Spatial--temporal analysis of multivariate environmental
monitoring data}.
In \bbooktitle{Multivariate Environmental Statistics}.
\bseries{North-Holland Ser. Statist. Probab.}
\bvolume{6}
\bpages{347--386}.
\bpublisher{North-Holland},
\blocation{Amsterdam}.
\bid{mr={1268443}}
\end{bincollection}
%
\bptok{imsref}%
\endbibitem

%b85 ###
\bibitem[\protect\citeauthoryear{Mearns et~al.}{2009}]{Meaetal09}
%
\begin{barticle}[auto:STB|2014/06/18|12:29:53]
\bauthor{\bsnm{Mearns},~\bfnm{L.~O.}\binits{L.~O.}},
\bauthor{\bsnm{Gutowski},~\bfnm{W.~J.}\binits{W.~J.}},
\bauthor{\bsnm{Jones},~\bfnm{R.}\binits{R.}},
\bauthor{\bsnm{Leung},~\bfnm{A.~M.}\binits{A.~M.}},
\bauthor{\bsnm{McGinnis},~\bfnm{B.}\binits{B.}},
\bauthor{\bsnm{Nunes},~\bfnm{Y.}\binits{Y.}} \AND
\bauthor{\bsnm{Qian},~\bfnm{Y.}\binits{Y.}}
(\byear{2009}).
\btitle{A regional climate change assessment program for North America}.
\bjournal{Eos, Transactions, American Geophysical Union}
\bvolume{90}
\bpages{311--312}.
\end{barticle}
%
\bptok{imsref}%
\endbibitem

%b86 ###
\bibitem[\protect\citeauthoryear{Myers}{1982}]{Mye82}
%
\begin{barticle}[mr]
\bauthor{\bsnm{Myers},~\bfnm{Donald~E.}\binits{D.~E.}}
(\byear{1982}).
\btitle{Matrix formulation of co-kriging}.
\bjournal{J. Internat. Assoc. Math. Geol.}
\bvolume{14}
\bpages{249--257}.
\bid{doi={10.1007/BF01032887}, issn={0020-5958}, mr={0660192}}
\end{barticle}
%
\bptok{imsref}%
% NOT OUTPUTED:
% issn = 0020-5958
% url = http://dx.doi.org/10.1007/BF01032887
% number = 3
% coden = IMGHBS
% fjournal = Journal of the International Association for Mathematical
%Geology
\endbibitem

%b87 ###
\bibitem[\protect\citeauthoryear{Myers}{1983}]{Mye83}
%
\begin{barticle}[mr]
\bauthor{\bsnm{Myers},~\bfnm{Donald~E.}\binits{D.~E.}}
(\byear{1983}).
\btitle{Estimation of linear combinations and co-{k}riging}.
\bjournal{J. Internat. Assoc. Math. Geol.}
\bvolume{15}
\bpages{633--637}.
\bid{doi={10.1007/BF01093416}, issn={0020-5958}, mr={0719602}}
\end{barticle}
%
\bptok{imsref}%
% NOT OUTPUTED:
% issn = 0020-5958
% url = http://dx.doi.org/10.1007/BF01093416
% number = 5
% coden = IMGHBS
% fjournal = Journal of the International Association for Mathematical
%Geology
\endbibitem

%b88 ###
\bibitem[\protect\citeauthoryear{Myers}{1991}]{Mye91}
%
\begin{barticle}[mr]
\bauthor{\bsnm{Myers},~\bfnm{Donald~E.}\binits{D.~E.}}
(\byear{1991}).
\btitle{Pseudo-cross variograms, positive-definiteness, and cokriging}.
\bjournal{Math. Geol.}
\bvolume{23}
\bpages{805--816}.
\bid{doi={10.1007/BF02068776}, issn={0882-8121}, mr={1121438}}
\end{barticle}
%
\bptok{imsref}%
% NOT OUTPUTED:
% issn = 0882-8121
% url = http://dx.doi.org/10.1007/BF02068776
% number = 6
% coden = MATGED
% fjournal = Mathematical Geology
\endbibitem

%b89 ###
\bibitem[\protect\citeauthoryear{Myers}{1992}]{Mye92}
%
\begin{barticle}[mr]
\bauthor{\bsnm{Myers},~\bfnm{Donald~E.}\binits{D.~E.}}
(\byear{1992}).
\btitle{Kriging, co-{k}riging, radial basis functions and the role of
positive definiteness}.
\bjournal{Comput. Math. Appl.}
\bvolume{24}
\bpages{139--148}.
%\bnote{Advances in the theory and applications of radial basis functions}.
\bid{doi={10.1016/0898-1221(92)90176-I}, issn={0898-1221}, mr={1190311}}
\end{barticle}
%
\bptok{imsref}%
% NOT OUTPUTED:
% issn = 0898-1221
% url = http://dx.doi.org/10.1016/0898-1221(92)90176-I
% number = 12
% coden = CMAPDK
% fjournal = Computers \& Mathematics with Applications. An
%International Journal
\endbibitem

%b90 ###
\bibitem[\protect\citeauthoryear{Narcowich and Ward}{1994}]{NarWar94}
%
\begin{barticle}[mr]
\bauthor{\bsnm{Narcowich},~\bfnm{Francis~J.}\binits{F.~J.}} \AND
\bauthor{\bsnm{Ward},~\bfnm{Joseph~D.}\binits{J.~D.}}
(\byear{1994}).
\btitle{Generalized {H}ermite interpolation via matrix-valued
conditionally positive definite functions}.
\bjournal{Math. Comp.}
\bvolume{63}
\bpages{661--687}.
\bid{doi={10.2307/2153288}, issn={0025-5718}, mr={1254147}}
\end{barticle}
%
\bptok{imsref}%
% NOT OUTPUTED:
% issn = 0025-5718
% url = http://dx.doi.org/10.2307/2153288
% number = 208
% coden = MCMPAF
% fjournal = Mathematics of Computation
\endbibitem

%b91 ###
\bibitem[\protect\citeauthoryear{North, Wang and Genton}{2011}]{NorWanGen11}
%
\begin{barticle}[auto:STB|2014/06/18|12:29:53]
\bauthor{\bsnm{North},~\bfnm{G.~R.}\binits{G.~R.}},
\bauthor{\bsnm{Wang},~\bfnm{J.}\binits{J.}} \AND
\bauthor{\bsnm{Genton},~\bfnm{M.~G.}\binits{M.~G.}}
(\byear{2011}).
\btitle{Correlation models for temperature fields}.
\bjournal{J. Climate}
\bvolume{24}
\bpages{5850--5862}.
\end{barticle}
%
\bptok{imsref}%
\endbibitem

%b92 ###
\bibitem[\protect\citeauthoryear{Oehlert}{1993}]{Oeh93}
%
\begin{barticle}[auto:STB|2014/06/18|12:29:53]
\bauthor{\bsnm{Oehlert},~\bfnm{G.~W.}\binits{G.~W.}}
(\byear{1993}).
\btitle{Regional trends in sulfate wet deposition}.
\bjournal{J. Amer. Statist. Assoc.}
\bvolume{88}
\bpages{390--399}.
\end{barticle}
%
\bptok{imsref}%
\endbibitem

%b93 ###
\bibitem[\protect\citeauthoryear{Paciorek and Schervish}{2006}]{PacSch06}
%
\begin{barticle}[mr]
\bauthor{\bsnm{Paciorek},~\bfnm{Christopher~J.}\binits{C.~J.}} \AND
\bauthor{\bsnm{Schervish},~\bfnm{Mark~J.}\binits{M.~J.}}
(\byear{2006}).
\btitle{Spatial modelling using a new class of nonstationary
covariance functions}.
\bjournal{Environmetrics}
\bvolume{17}
\bpages{483--506}.
\bid{doi={10.1002/env.785}, issn={1180-4009}, mr={2240939}}
\end{barticle}
%
\bptok{imsref}%
% NOT OUTPUTED:
% issn = 1180-4009
% url = http://dx.doi.org/10.1002/env.785
% number = 5
% fjournal = Environmetrics
\endbibitem

%b94 ###
\bibitem[\protect\citeauthoryear{Papritz, K{\"u}nsch and
Webster}{1993}]{PapKunWeb93}
%
\begin{barticle}[mr]
\bauthor{\bsnm{Papritz},~\bfnm{A.}\binits{A.}},
\bauthor{\bsnm{K{\"u}nsch},~\bfnm{H.~R.}\binits{H.~R.}} \AND
\bauthor{\bsnm{Webster},~\bfnm{R.}\binits{R.}}
(\byear{1993}).
\btitle{On the pseudo cross-variogram}.
\bjournal{Math. Geol.}
\bvolume{25}
\bpages{1015--1026}.
\bid{doi={10.1007/BF00911547}, issn={0882-8121}, mr={1246957}}
\end{barticle}
%
\bptok{imsref}%
% NOT OUTPUTED:
% issn = 0882-8121
% url = http://dx.doi.org/10.1007/BF00911547
% number = 8
% coden = MATGED
% fjournal = Mathematical Geology
\endbibitem

%b95 ###
\bibitem[\protect\citeauthoryear{Pascual and Zhang}{2006}]{PasZha06}
%
\begin{barticle}[mr]
\bauthor{\bsnm{Pascual},~\bfnm{F.}\binits{F.}} \AND
\bauthor{\bsnm{Zhang},~\bfnm{H.}\binits{H.}}
(\byear{2006}).
\btitle{Estimation of linear correlation coefficient of two correlated
spatial processes}.
\bjournal{Sankhy\=a}
\bvolume{68}
\bpages{307--325}.
\bid{issn={0972-7671}, mr={2303086}}
\end{barticle}
%
\bptok{imsref}%
% NOT OUTPUTED:
% issn = 0972-7671
% number = 2
% fjournal = Sankhy\=a. The Indian Journal of Statistics
\endbibitem

%b96 ###
\bibitem[\protect\citeauthoryear{Peterson and Vose}{1997}]{PetVos97}
%
\begin{barticle}[auto:STB|2014/06/18|12:29:53]
\bauthor{\bsnm{Peterson},~\bfnm{T.~C.}\binits{T.~C.}} \AND
\bauthor{\bsnm{Vose},~\bfnm{R.~S.}\binits{R.~S.}}
(\byear{1997}).
\btitle{An overview of the Global Historical Climatology Network
temperature database}.
\bjournal{Bull. Amer. Meteorol. Soc.}
\bvolume{78}
\bpages{2837--2849}.
\end{barticle}
%
\bptok{imsref}%
\endbibitem

%b97 ###
\bibitem[\protect\citeauthoryear{Porcu, Bevilacqua and
Genton}{2014}]{PorBevGen}
%
\begin{bmisc}[auto:STB|2014/06/18|12:29:53]
\bauthor{\bsnm{Porcu},~\bfnm{E.}\binits{E.}},
\bauthor{\bsnm{Bevilacqua},~\bfnm{M.}\binits{M.}} \AND
\bauthor{\bsnm{Genton},~\bfnm{M.~G.}\binits{M.~G.}}
(\byear{2014}).
\bhowpublished{Spatio-temporal covariance and cross-covariance
functions of the great circle distance on a sphere. Unpublished manuscript}.
\end{bmisc}
%
\bptok{imsref}%
% NOT OUTPUTED:
% sortkey = Porcu(2014
% howpublished = (2014), ``Spatio-temporal covariance and
%cross-covariance functions of the great circle distance on a sphere, "
%unpublished manuscript
\endbibitem

%b98 ###
\bibitem[\protect\citeauthoryear{Porcu, Gregori and
Mateu}{2006}]{PorGreMat06}
%
\begin{barticle}[mr]
\bauthor{\bsnm{Porcu},~\bfnm{E.}\binits{E.}},
\bauthor{\bsnm{Gregori},~\bfnm{P.}\binits{P.}} \AND
\bauthor{\bsnm{Mateu},~\bfnm{J.}\binits{J.}}
(\byear{2006}).
\btitle{Nonseparable stationary anisotropic space--time covariance functions}.
\bjournal{Stoch. Environ. Res. Risk Assess.}
\bvolume{21}
\bpages{113--122}.
\bid{doi={10.1007/s00477-006-0048-3}, issn={1436-3240}, mr={2307631}}
\end{barticle}
%
\bptok{imsref}%
% NOT OUTPUTED:
% issn = 1436-3240
% url = http://dx.doi.org/10.1007/s00477-006-0048-3
% number = 2
% fjournal = Stochastic Environmental Research and Risk Assessment
%(SERRA)
\endbibitem

%b99 ###
\bibitem[\protect\citeauthoryear{Porcu and Zastavnyi}{2011}]{PorZas11}
%
\begin{barticle}[mr]
\bauthor{\bsnm{Porcu},~\bfnm{Emilio}\binits{E.}} \AND
\bauthor{\bsnm{Zastavnyi},~\bfnm{Viktor}\binits{V.}}
(\byear{2011}).
\btitle{Characterization theorems for some classes of covariance
functions associated to vector valued random fields}.
\bjournal{J. Multivariate Anal.}
\bvolume{102}
\bpages{1293--1301}.
\bid{doi={10.1016/j.jmva.2011.04.013}, issn={0047-259X}, mr={2811618}}
\end{barticle}
%
\bptok{imsref}%
% NOT OUTPUTED:
% issn = 0047-259X
% url = http://dx.doi.org/10.1016/j.jmva.2011.04.013
% number = 9
% fjournal = Journal of Multivariate Analysis
\endbibitem

%b100 ###
\bibitem[\protect\citeauthoryear{Porcu et~al.}{2013}]{Poretal13}
%
\begin{barticle}[auto:STB|2014/06/18|12:29:53]
\bauthor{\bsnm{Porcu},~\bfnm{E.}\binits{E.}},
\bauthor{\bsnm{Daley},~\bfnm{D.~J.}\binits{D.~J.}},
\bauthor{\bsnm{Buhmann},~\bfnm{M.}\binits{M.}} \AND
\bauthor{\bsnm{Bevilacqua},~\bfnm{M.}\binits{M.}}
(\byear{2013}).
\btitle{Radial basis functions with compact support for multivariate
geostatistics}.
\bjournal{Stoch. Environ. Res. Risk Assess.}
\bvolume{27}
\bpages{909--922}.
\end{barticle}
%
\bptok{imsref}%
\endbibitem

%b101 ###
\bibitem[\protect\citeauthoryear{Reisert and Burk\-hardt}{2007}]{ReiBur07}
%
\begin{barticle}[mr]
\bauthor{\bsnm{Reisert},~\bfnm{Marco}\binits{M.}} \AND
\bauthor{\bsnm{Burkhardt},~\bfnm{Hans}\binits{H.}}
(\byear{2007}).
\btitle{Learning equivariant functions with matrix valued kernels}.
\bjournal{J. Mach. Learn. Res.}
\bvolume{8}
\bpages{385--408}.
\bid{issn={1532-4435}, mr={2320676}}
\end{barticle}
%
\bptok{imsref}%
% NOT OUTPUTED:
% issn = 1532-4435
% fjournal = Journal of Machine Learning Research (JMLR)
\endbibitem

%b102 ###
\bibitem[\protect\citeauthoryear{Rouhani and Wackernagel}{1990}]{RouWac90}
%
\begin{barticle}[auto:STB|2014/06/18|12:29:53]
\bauthor{\bsnm{Rouhani},~\bfnm{S.}\binits{S.}} \AND
\bauthor{\bsnm{Wackernagel},~\bfnm{H.}\binits{H.}}
(\byear{1990}).
\btitle{Multivariate geostatistical approach to space--time data analysis}.
\bjournal{Water Resour. Res.}
\bvolume{26}
\bpages{585--591}.
\end{barticle}
%
\bptok{imsref}%
\endbibitem

%b103 ###
\bibitem[\protect\citeauthoryear{Royle and Berliner}{1999}]{RoyBer99}
%
\begin{barticle}[mr]
\bauthor{\bsnm{Royle},~\bfnm{J.~Andrew}\binits{J.~A.}} \AND
\bauthor{\bsnm{Berliner},~\bfnm{L.~Mark}\binits{L.~M.}}
(\byear{1999}).
\btitle{A hierarchical approach to multivariate spatial modeling and
prediction}.
\bjournal{J. Agric. Biol. Environ. Stat.}
\bvolume{4}
\bpages{29--56}.
\bid{doi={10.2307/1400420}, issn={1085-7117}, mr={1812239}}
\end{barticle}
%
\bptok{imsref}%
% NOT OUTPUTED:
% issn = 1085-7117
% url = http://dx.doi.org/10.2307/1400420
% number = 1
% fjournal = Journal of Agricultural, Biological, and Environmental
%Statistics
\endbibitem

%b104 ###
\bibitem[\protect\citeauthoryear{Ruiz-Medina and Porcu}{2015}]{RuiPor}
%
\begin{barticle}[auto:STB|2014/06/18|12:29:53]
\bauthor{\bsnm{Ruiz-Medina},~\bfnm{M.~D.}\binits{M.~D.}} \AND
\bauthor{\bsnm{Porcu},~\bfnm{E.}\binits{E.}}
(\byear{2015}).
\btitle{Equivalence of Gaussian measures of multivariate random
fields}.
\bjournal{Stoch. Environ. Res. Risk Assess.}
\bvolume{29}
\bpages{325--334}.
\end{barticle}
%
\bptok{imsref}%
% NOT OUTPUTED:
% sortkey = Ruiz(2013
% howpublished = (2013), ``Equivalence of Gaussian measures of
%multivariate random fields, " \textit{Technical Report University
%Federico Santa Maria}
\endbibitem

%b105 ###
\bibitem[\protect\citeauthoryear{Sain and Cressie}{2007}]{SaiCre07}
%
\begin{barticle}[mr]
\bauthor{\bsnm{Sain},~\bfnm{Stephan~R.}\binits{S.~R.}} \AND
\bauthor{\bsnm{Cressie},~\bfnm{Noel}\binits{N.}}
(\byear{2007}).
\btitle{A spatial model for multivariate lattice data}.
\bjournal{J. Econometrics}
\bvolume{140}
\bpages{226--259}.
\bid{doi={10.1016/j.jeconom.2006.09.010}, issn={0304-4076}, mr={2395923}}
\end{barticle}
%
\bptok{imsref}%
% NOT OUTPUTED:
% issn = 0304-4076
% url = http://dx.doi.org/10.1016/j.jeconom.2006.09.010
% number = 1
% coden = JECMB6
% fjournal = Journal of Econometrics
\endbibitem

%b106 ###
\bibitem[\protect\citeauthoryear{Sampson and Guttorp}{1992}]{SamGut92}
%
\begin{barticle}[auto:STB|2014/06/18|12:29:53]
\bauthor{\bsnm{Sampson},~\bfnm{P.~D.}\binits{P.~D.}} \AND
\bauthor{\bsnm{Guttorp},~\bfnm{P.}\binits{P.}}
(\byear{1992}).
\btitle{Nonparametric estimation of nonstationary spatial covariance
structure}.
\bjournal{J. Amer. Statist. Assoc.}
\bvolume{87}
\bpages{108--119}.
\end{barticle}
%
\bptok{imsref}%
\endbibitem

%b107 ###
\bibitem[\protect\citeauthoryear{Sang, Jun and Huang}{2011}]{SanJunHua11}
%
\begin{barticle}[mr]
\bauthor{\bsnm{Sang},~\bfnm{Huiyan}\binits{H.}},
\bauthor{\bsnm{Jun},~\bfnm{Mikyoung}\binits{M.}} \AND
\bauthor{\bsnm{Huang},~\bfnm{Jianhua~Z.}\binits{J.~Z.}}
(\byear{2011}).
\btitle{Covariance approximation for large multivariate spatial data
sets with an application to multiple climate model errors}.
\bjournal{Ann. Appl. Stat.}
\bvolume{5}
\bpages{2519--2548}.
\bid{doi={10.1214/11-AOAS478}, issn={1932-6157}, mr={2907125}}
\end{barticle}
%
\bptok{imsref}%
% NOT OUTPUTED:
% issn = 1932-6157
% url = http://dx.doi.org/10.1214/11-AOAS478
% number = 4
% fjournal = The Annals of Applied Statistics
\endbibitem

%b108 ###
\bibitem[\protect\citeauthoryear{Scheuerer and Schlather}{2012}]{SchSch12}
%
\begin{barticle}[mr]
\bauthor{\bsnm{Scheuerer},~\bfnm{Michael}\binits{M.}} \AND
\bauthor{\bsnm{Schlather},~\bfnm{Martin}\binits{M.}}
(\byear{2012}).
\btitle{Covariance models for divergence-free and curl-free random
vector fields}.
\bjournal{Stoch. Models}
\bvolume{28}
\bpages{433--451}.
\bid{doi={10.1080/15326349.2012.699756}, issn={1532-6349}, mr={2959449}}
\end{barticle}
%
\bptok{imsref}%
% NOT OUTPUTED:
% issn = 1532-6349
% url = http://dx.doi.org/10.1080/15326349.2012.699756
% number = 3
% coden = CSSME8
% fjournal = Stochastic Models
\endbibitem

%b109 ###
\bibitem[\protect\citeauthoryear{Schlather}{2010}]{Sch10}
%
\begin{barticle}[mr]
\bauthor{\bsnm{Schlather},~\bfnm{Martin}\binits{M.}}
(\byear{2010}).
\btitle{Some covariance models based on normal scale mixtures}.
\bjournal{Bernoulli}
\bvolume{16}
\bpages{780--797}.
\bid{doi={10.3150/09-BEJ226}, issn={1350-7265}, mr={2730648}}
\end{barticle}
%
\bptok{imsref}%
% NOT OUTPUTED:
% issn = 1350-7265
% url = http://dx.doi.org/10.3150/09-BEJ226
% number = 3
% fjournal = Bernoulli. Official Journal of the Bernoulli Society for
%Mathematical Statistics and Probability
\endbibitem

%b110 ###
\bibitem[\protect\citeauthoryear{Schmidt and Gelfand}{2003}]{SchGel}
%
\begin{barticle}[auto:STB|2014/06/18|12:29:53]
\bauthor{\bsnm{Schmidt},~\bfnm{A.~M.}\binits{A.~M.}} \AND
\bauthor{\bsnm{Gelfand},~\bfnm{A.~E.}\binits{A.~E.}}
(\byear{2003}).
\btitle{A Bayesian coregionalization approach for multivariate
pollutant data}.
\bjournal{J. Geophys. Res.}
\bvolume{108}
%\bnote{STS 10}
\bpages{1--9}.
\end{barticle}
%
\bptok{imsref}%
% NOT OUTPUTED:
% sortkey = Schmidt(2003
% howpublished = (2003), ``A Bayesian coregionalization approach for
%multivariate pollutant data, ''
% \textit{Journal of Geophysical Research}, 108, STS 10, 1-9
\endbibitem

%b111 ###
\bibitem[\protect\citeauthoryear{Shmaryan and Journel}{1999}]{ShmJou99}
%
\begin{barticle}[mr]
\bauthor{\bsnm{Shmaryan},~\bfnm{L.~E.}\binits{L.~E.}} \AND
\bauthor{\bsnm{Journel},~\bfnm{A.~G.}\binits{A.~G.}}
(\byear{1999}).
\btitle{Two {M}arkov models and their application}.
\bjournal{Math. Geol.}
\bvolume{31}
\bpages{965--988}.
\bid{doi={10.1023/A:1007505130226}, issn={0882-8121}, mr={1717034}}
\end{barticle}
%
\bptok{imsref}%
% NOT OUTPUTED:
% issn = 0882-8121
% url = http://dx.doi.org/10.1023/A:1007505130226
% number = 8
% coden = MATGED
% fjournal = Mathematical Geology
\endbibitem

%b112 ###
\bibitem[\protect\citeauthoryear{Stein}{1999}]{Ste99}
%
\begin{bbook}[mr]
\bauthor{\bsnm{Stein},~\bfnm{Michael~L.}\binits{M.~L.}}
(\byear{1999}).
\btitle{Interpolation of Spatial Data: Some Theory for Kriging}.
\bseries{Springer Series in Statistics}.
\bpublisher{Springer},
\blocation{New York}.
\bid{doi={10.1007/978-1-4612-1494-6}, mr={1697409}}
\end{bbook}
%
\bptok{imsref}%
% NOT OUTPUTED:
% isbn = 0-387-98629-4
% url = http://dx.doi.org/10.1007/978-1-4612-1494-6
% fpage = xviii+247
\endbibitem

%b113 ###
\bibitem[\protect\citeauthoryear{Stein}{2005}]{Ste05}
%
\begin{bmisc}[auto:STB|2014/06/18|12:29:53]
\bauthor{\bsnm{Stein},~\bfnm{M.~L.}\binits{M.~L.}}
(\byear{2005}).
\bhowpublished{Nonstationary spatial covariance functions.
Technical Report 21, Univ. Chicago, CISES}.
\end{bmisc}
%
\bptok{imsref}%
\endbibitem

%b114 ###
\bibitem[\protect\citeauthoryear{Subramanyam and Pandalai}{2008}]{SubPan08}
%
\begin{barticle}[mr]
\bauthor{\bsnm{Subramanyam},~\bfnm{A.}\binits{A.}} \AND
\bauthor{\bsnm{Pandalai},~\bfnm{H.~S.}\binits{H.~S.}}
(\byear{2008}).
\btitle{Data configurations and the cokriging system: Simplification
by screen effects}.
\bjournal{Math. Geosci.}
\bvolume{40}
\bpages{425--443}.
\bid{doi={10.1007/s11004-008-9153-9}, issn={1874-8961}, mr={2401515}}
\end{barticle}
%
\bptok{imsref}%
% NOT OUTPUTED:
% issn = 1874-8961
% url = http://dx.doi.org/10.1007/s11004-008-9153-9
% number = 4
% fjournal = Mathematical Geosciences
\endbibitem

%b115 ###
\bibitem[\protect\citeauthoryear{Sun, Li and Genton}{2012}]{SunLiGen}
%
\begin{bincollection}[auto:STB|2014/06/18|12:29:53]
\bauthor{\bsnm{Sun},~\bfnm{Y.}\binits{Y.}},
\bauthor{\bsnm{Li},~\bfnm{B.}\binits{B.}} \AND
\bauthor{\bsnm{Genton},~\bfnm{M.~G.}\binits{M.~G.}}
(\byear{2012}).
\btitle{Geostatistics for large datasets}.
In \bbooktitle{Space--Time Processes and Challenges Related to
Environmental Problems}
\bvolume{207}
(\beditor{\bfnm{E.}\binits{E.} \bsnm{Porcu}},
\beditor{\bfnm{J. M.}\binits{J. M.} \bsnm{Montero}} \AND
\beditor{\bfnm{M.}\binits{M.} \bsnm{Schlather}}, eds.)
\bpages{55--77}.
\bpublisher{Springer},
\blocation{Berlin}.
%\bnote{Chapter~3}.
\end{bincollection}
%
\bptok{imsref}%
% NOT OUTPUTED:
% sortkey = Sun(2012
% howpublished = (2012), ``Geostatistics for large datasets, " in
%\textit{Space-Time Processes and Challenges Related to Environmental
%Problems}, E. Porcu, J. M. Montero, M. Schlather (eds), Springer, Vol.
%207, Chapter 3, 55-77
\endbibitem

%b116 ###
\bibitem[\protect\citeauthoryear{Vargas-Guzm{\'a}n, Warrick and
Myers}{1999}]{VarWarMye99}
%
\begin{barticle}[auto:STB|2014/06/18|12:29:53]
\bauthor{\bsnm{Vargas-Guzm{\'a}n},~\bfnm{J.~A.}\binits{J.~A.}},
\bauthor{\bsnm{Warrick},~\bfnm{A.~W.}\binits{A.~W.}} \AND
\bauthor{\bsnm{Myers},~\bfnm{D.~E.}\binits{D.~E.}}
(\byear{1999}).
\btitle{Multivariate correlation in the framework of support and
spatial scales of variability}.
\bjournal{Math. Geol.}
\bvolume{31}
\bpages{85--104}.
\end{barticle}
%
\bptok{imsref}%
\endbibitem

%b117 ###
\bibitem[\protect\citeauthoryear{Vargas-Guzm{\'a}n, Warrick and
Myers}{2002}]{VarWarMye02}
%
\begin{barticle}[mr]
\bauthor{\bsnm{Vargas-Guzm{\'a}n},~\bfnm{J.~A.}\binits{J.~A.}},
\bauthor{\bsnm{Warrick},~\bfnm{A.~W.}\binits{A.~W.}} \AND
\bauthor{\bsnm{Myers},~\bfnm{D.~E.}\binits{D.~E.}}
(\byear{2002}).
\btitle{Coregionalization by linear combination of nonorthogonal components}.
\bjournal{Math. Geol.}
\bvolume{34}
\bpages{405--419}.
\bid{doi={10.1023/A:1015078911063}, issn={0882-8121}, mr={1951789}}
\end{barticle}
%
\bptok{imsref}%
% NOT OUTPUTED:
% issn = 0882-8121
% url = http://dx.doi.org/10.1023/A:1015078911063
% number = 4
% coden = MATGED
% fjournal = Mathematical Geology
\endbibitem

%b118 ###
\bibitem[\protect\citeauthoryear{Ver~Hoef and Barry}{1998}]{VerBar98}
%
\begin{barticle}[mr]
\bauthor{\bsnm{Ver Hoef},~\bfnm{Jay~M.}\binits{J.~M.}} \AND
\bauthor{\bsnm{Barry},~\bfnm{Ronald~Paul}\binits{R.~P.}}
(\byear{1998}).
\btitle{Constructing and fitting models for cokriging and
multivariable spatial prediction}.
\bjournal{J. Statist. Plann. Inference}
\bvolume{69}
\bpages{275--294}.
\bid{doi={10.1016/S0378-3758(97)00162-6}, issn={0378-3758}, mr={1631328}}
\end{barticle}
%
\bptok{imsref}%
% NOT OUTPUTED:
% issn = 0378-3758
% url = http://dx.doi.org/10.1016/S0378-3758(97)00162-6
% number = 2
% coden = JSPIDN
% fjournal = Journal of Statistical Planning and Inference
\endbibitem

%b119 ###
\bibitem[\protect\citeauthoryear{Ver~Hoef and Cressie}{1993}]{VerCre93}
%
\begin{barticle}[mr]
\bauthor{\bsnm{Ver Hoef},~\bfnm{Jay~M.}\binits{J.~M.}} \AND
\bauthor{\bsnm{Cressie},~\bfnm{Noel}\binits{N.}}
(\byear{1993}).
\btitle{Multivariable spatial prediction}.
\bjournal{Math. Geol.}
\bvolume{25}
\bpages{219--240}.
\bid{doi={10.1007/BF00893273}, issn={0882-8121}, mr={1206187}}
\end{barticle}
%
\bptok{imsref}%
% NOT OUTPUTED:
% issn = 0882-8121
% url = http://dx.doi.org/10.1007/BF00893273
% number = 2
% coden = MATGED
% fjournal = Mathematical Geology
\endbibitem

%b120 ###
\bibitem[\protect\citeauthoryear{Ver~Hoef, Cressie and
Barry}{2004}]{VerCreBar04}
%
\begin{barticle}[mr]
\bauthor{\bsnm{Ver Hoef},~\bfnm{Jay~M.}\binits{J.~M.}},
\bauthor{\bsnm{Cressie},~\bfnm{Noel}\binits{N.}} \AND
\bauthor{\bsnm{Barry},~\bfnm{Ronald~Paul}\binits{R.~P.}}
(\byear{2004}).
\btitle{Flexible spatial models for kriging and cokriging using moving
averages and the fast {F}ourier transform (FFT)}.
\bjournal{J. Comput. Graph. Statist.}
\bvolume{13}
\bpages{265--282}.
\bid{doi={10.1198/1061860043498}, issn={1061-8600}, mr={2063985}}
\end{barticle}
%
\bptok{imsref}%
% NOT OUTPUTED:
% issn = 1061-8600
% url = http://dx.doi.org/10.1198/1061860043498
% number = 2
% fjournal = Journal of Computational and Graphical Statistics
\endbibitem

%b121 ###
\bibitem[\protect\citeauthoryear{Wackernagel}{1994}]{Wac94}
%
\begin{barticle}[auto:STB|2014/06/18|12:29:53]
\bauthor{\bsnm{Wackernagel},~\bfnm{H.}\binits{H.}}
(\byear{1994}).
\btitle{Cokriging versus kriging in regionalized multivariate data analysis}.
\bjournal{Geoderma}
\bvolume{62}
\bpages{83--92}.
\end{barticle}
%
\bptok{imsref}%
\endbibitem

%b122 ###
\bibitem[\protect\citeauthoryear{Wackernagel}{2003}]{kse03}
%
\begin{bbook}[auto:STB|2014/06/18|12:29:53]
\bauthor{\bsnm{Wackernagel},~\bfnm{H.}\binits{H.}}
(\byear{2003}).
\btitle{Multivariate Geostatistics: An Introduction with
Applications},
\bedition{3rd} ed.
\bpublisher{Springer},
\blocation{Berlin}.
\end{bbook}
%
\bptok{imsref}%
\endbibitem

%b123 ###
\bibitem[\protect\citeauthoryear{Zhang}{2007}]{Zha07}
%
\begin{barticle}[mr]
\bauthor{\bsnm{Zhang},~\bfnm{Hao}\binits{H.}}
(\byear{2007}).
\btitle{Maximum-likelihood estimation for multivariate spatial linear
coregionalization models}.
\bjournal{Environmetrics}
\bvolume{18}
\bpages{125--139}.
\bid{doi={10.1002/env.807}, issn={1180-4009}, mr={2345650}}
\end{barticle}
%
\bptok{imsref}%
% NOT OUTPUTED:
% issn = 1180-4009
% url = http://dx.doi.org/10.1002/env.807
% number = 2
% fjournal = Environmetrics
\endbibitem

\end{thebibliography}
\end{document}